\documentclass[preprint]{aastex}

\newcommand{\etal}{{\it et~al.}}
\newcommand{\sqasec}{\sq\arcsec}
\newcommand{\HI}{H{\tt I}}

\newcommand{\kms}{km\thinspace s$^{-1}$}
\newcommand{\solar}{_\odot}
\newcommand{\Ha}{H$\alpha$}

\slugcomment{To Be Published by The Astronomical Journal (July 1, 2002)}

\shorttitle{Galaxies on the Blue Edge}
\shortauthors{Cabanela and Dickey}

\begin{document}


\title{Galaxies on the Blue Edge}


\author{J. E. Cabanela \altaffilmark{1}}
\affil{Haverford College, 370 Lancaster Ave, Haverford, PA 19041}
\email{jcabanel@haverford.edu}

\and

\author{J. M. Dickey}
\affil{Astronomy Department, University of Minnesota \\ 116 Church Street SE, Minneapolis, MN 55455}
\email{john@astro.umn.edu}



\begin{abstract}
We have successfully constructed a catalog of {\HI}-rich galaxies selected from
the Minnesota Automated Plate Scanner Catalog of the Palomar Observatory
Sky Survey (POSS I) based solely on optical criteria.   We identify {\HI}-rich
candidates by selecting the bluest galaxies at a given apparent magnitude, those
galaxies on the blue edge of POSS I color-magnitude parameter space.
Subsequent 21-cm observations on the upgraded Arecibo 305m dish detected over
50\% of the observed candidates.  The detected galaxies are {\HI}-rich with
{\HI} masses comparable to ``normal'' high surface brightness disk galaxies and
they have gas mass-to-light ratios ranging from 0.1 to 4.8 (in solar units).
Comparison of our candidate galaxies with known low surface brightness galaxies
(hereafter LSBs) shows that they exhibit similar optical and {\HI} properties to
that population.  We also show that previously identified LSBs, including
several LSBs with red $B-V$ colors, preferentially occupy the blue edge of
POSS I color-magnitude parameter space.  Their presence on the blue edge
appears to be a selection effect due to differing plate limits in the two POSS I
bandpasses.  This suggests the POSS I is a good filter for separating galaxies
on the higher surface brightness end of the LSB population from the general
population of galaxies in the night sky.
\end{abstract}







\section{Introduction}

Low surface brightness galaxies (LSBs) are an important species of extragalactic
object, differing from ``normal'' high surface brightness (HSB) galaxies in that
their stellar disks are more diffuse (see Bothun, Impey, and McGaugh 1997).
Their presence went largely unnoticed until the 1980s because they have central
surface brightnesses lower than the sky surface brightness ($\mu_{sky} \approx
23$ mag/{\sqasec} in V). \citet{dis76} demonstrated that most earlier galaxy
catalogs had limiting surface brightnesses near the sky surface brightness and
speculated that ``what is seen above the sky background may be no reliable
measure of what lies underneath.'' This speculation proved correct in 1987 with
the discovery of Malin 1 \citep{bot87}, which appears as a small dwarf in short
exposures on photographic plates (such as those of the Palomar Observatory Sky
Survey [POSS I]), but reveals itself to be the largest known disk galaxy in deep
exposures.  Since then, many deep surveys have been conducted to investigate
this previously unexplored population of LSBs.

A proposed explanation for the low surface brightness nature of LSBs is that
they are probably very slowly evolving systems and have very low current star
formation rates. This hypothesis is supported by the fact that while LSBs are
typically {\HI}-rich, their gas surface densities are observed to lie below the
Toomre threshold for star formation over most of their gas disks \citep{vdH93,
dBl96, vZe99}. The Toomre threshold density varies across a galaxy, and thus it
is possible that while their densities are globally subcritical, there are
regions of LSBs that temporarily meet the Toomre criterion for star formation,
and thus some star formation occurs. \citet{mar01} cite examples of globally
subcritical galaxies in which star formation is actually quite vigorous (e.g.
NGC 2403 and M33). This theory could also explain why most LSBs are very blue,
since most of the luminosity in these galaxies comes from a relatively few young
stars, although the low metallicity of LSBs also appears to play a role
\citep{ger99}.

LSBs present a testbed for probing the evolution of stars and the interstellar
medium (ISM) in low gas density environments. For example, it has been argued
that Blue Compact Dwarfs (BCDs), {\HI}-rich galaxies experiencing a current
burst of star formation, are simply the cores of low surface brightness extended
disks \citep{meu96}.  This potential relationship between BCDs and LSBs is
contested \citep{sal99}, but it points out that we are still determining ``what
lies underneath'' the sky brightness, so a large sample size is important to
developing an understanding of the physical nature of these galaxies.  There is
thus strong motivation to construct a large catalog of LSBs and similar
{\HI}-rich galaxies. 

One approach to finding these {\HI}-rich galaxies has been blind radio surveys
at $\lambda$21-cm \citep{sch96,bri97,dic97,sta01}, which have the distinct
advantage of not being affected by optical selection effects.  However, blind
radio surveys may be considered by some an inefficient use of resources due to
the large amount of telescope time needed per detection obtained.   Thus,
despite the obvious selection effects, until recently, most {\HI}-rich LSB
galaxies were discovered through searches of wide-field optical surveys,
typically involving the visual examination of photographic plates to identify
objects with low central surface brightness (See \citet{sch92}, \citet{spr96},
and \citet{imp97} and the surveys cited therein). Followup is then performed at
$\lambda$21-cm in order to confirm a high {\HI} content. Some semi-automated
searches of photographic plates have also been performed. For example,
\citet{imp96} used a combination of multiple optical parameters (including
surface brightness) with visual followup in order to identify LSB candidates
from the APM scans of UK Schmidt photographic plates.  One limitation of
searching photographic plates is that plate-based surveys can say nothing about
the most diffuse LSBs, those below the plate limit.  This problem was addressed
in the CCD survey of \citet{one97a} which has lower $B$ surface brightness
limits than photographic surveys, although a complete visual inspection of the
images was necessary to identify LSB candidates.  An automated mechanism for
identifying LSBs or other {\HI}-rich, low gas density galaxy candidates from
wide-field optical surveys could certainly speed up the construction of a large
catalog of LSBs. In addition, the automation could avoid the need for difficult
to quantify subjective selection criteria for catalog members.

\section{{\HI}-rich Galaxies in Color-Magnitude Parameter Space \label{HercReview}}

In a recent study of Hercules Cluster galaxies \citep{dic97}, we performed an
$\lambda$21-cm flux-limited survey of galaxies in four VLA fields (designated
``NE,'' ``CE,'', ``SW,'' and ``47'') overlaying the Hercules cluster using the
Very Large Array. The impetus for this initial study was to examine the {\HI}
properties of galaxies in the variety of environments present throughout the
Hercules cluster. Motivated by an interest in possible optical counterparts to
these {\HI}-selected galaxies, we cross-identified the \citet{dic97} {\HI}
catalog (hereafter the Hercules {\HI} catalog) with an optical catalog from the
Minnesota Automated Plate Scanner Catalog of the POSS I (\citet{ode92,pen93},
hereafter the APS Catalog).  The APS Catalog is constructed from digitized scans
of both the blue 103a-O (hereafter O) and red 103-E (hereafter E) plates of the
POSS I.  The APS Catalog can be distinguished
from other digitized versions of the POSS I (such as the STScI Digitized Sky
Survey) in part by the fact that each POSS I field was independently
photometrically calibrated and both plates were scanned in each field, providing
$O-E$ color information.

We obtained optical cross-identifications for 51 of the Hercules {\HI} galaxies
in the APS Catalog.  For comparison, we also retrieved all the galaxies from the
APS catalog in the four VLA fields covered in the Hercules {\HI} study (a total of
7971 galaxies).  All the Hercules catalog regions lie in a single POSS I field
(P445\footnote{The POSS I field designations we use are the modified
Luyten POSS I plate numbers, which are typically one greater than the plate
reference numbers used in the Guide Star Catalog \citep{las90}.}), so
plate-to-plate photometry zero-point variations do not affect the relative
colors of these galaxies.  We constructed a color-magnitude diagram using $O-E$
and $E$ magnitude data for both the Hercules {\HI} catalog and all the APS
galaxies in the same areas (shown in Figure \ref{colormag}).  It is evident from
this diagram that the Hercules {\HI} catalog galaxies are bluer than the
majority of galaxies in the range $16 <\ m_{O} < 21$. Therefore most of the
optical counterparts for this {\HI}-selected sample of galaxies are on the
blue edge of the color-magnitude diagram.

To more precisely quantify this effect, we define the blue edge boundary as the
$O-E$ color separating the bluest 10\% of galaxies of comparable $E$ magnitude
($\pm 0.25^m$) in POSS I field P445 from the rest of the galaxies in that field.
At brighter magnitudes, there are too few galaxies to define the blue
edge in this manner, so we instead require the blue edge boundary to lie 0.3
magnitude redward of bluest galaxy. This boundary is fit with a 2$^{nd}$
order polynomial, such that
\begin{equation}
(O-E)_{BE} = \left\{
\begin{array}{r l}
 -10.172 + 1.4984 m_E - 0.048391 m_E^2 \\ 
 1.427 
 \end{array}\right\} \mbox{where} \left\{
\begin{array}{r l}
m_E \geq 15.48 \\ m_E < 15.48
\end{array}\right. 
\label{BEeqn}
\end{equation}
where $m_E$ is the extinction-corrected $E$ magnitude of the
galaxy.\footnote{Due to variations in the photometry zeropoint from plate to
plate, this equation for the blue edge is technically only valid for P445,
although it should be a good estimate of the location of the blue edge
boundary for most POSS I fields.} We determined $A_O$ and $A_E$ for each galaxy
in our sample using the E(B-V) maps from \citet{sch98} in conjunction with the
UV to near-IR extinction law from \citet{car89}.  We also tested extinction
corrections based on the E(B-V) maps from \citet{bur82} and found no substantial
differences in our results.  

The blue edge boundaries plotted on all the color-magnitude diagrams in this
paper are defined by Equation \ref{BEeqn}. Using this definition of the ``blue
edge'' boundary, we find over 80\% of the Hercules {\HI} galaxies lie blueward
this blue edge boundary.  Therefore, we speculated that this might be a
property common to {\HI}-rich galaxies, including the {\HI}-rich LSBs.
Consequently, an effective strategy for automatic optical selection of
{\HI}-rich candidates for an {\HI} survey could be to simply observe the
galaxies on the blue edge of POSS I color-magnitude parameter space. 

\section{Probing the blue edge of the Pisces-Perseus Supercluster
\label{probing}}

To test this hypothesis that blue edge galaxies are in general also
{\HI}-rich, we extracted a sample of such blue edge galaxies from the APS
Catalog in the POSS I fields covering the Perseus-Pisces Supercluster.  The
Pisces-Perseus Supercluster (hereafter PPS) is one of the largest structures in
the nearby universe.  At a redshift between 4000 and 6000 {\kms}, the PPS and
its surrounding filaments cover over 4000 square degrees of the sky, implying a
linear extent of approximately 30$h^{-1}$ by 50$h^{-1}$ Mpc! The densest portion
of the PPS, the ``ridgeline'' as defined by \citet{cab98}, extends from
($\alpha,\delta$) of (22$^h$,40{\arcdeg}) through (23$^h$,28{\arcdeg}) to
(3$^h$,40{\arcdeg}) (the region near the galaxy cluster A426).  The extent of
this supercluster is illustrated in Figure 1 of \citet{gio86} where it can also
be seen that the PPS encompasses a very broad range of density environments,
ranging from the high density regions of several rich clusters (including A262,
A347, and A426) through a large volume with much lower densities, but higher
than the field.

As noted by \citet{gio85}, the {\HI} content of a galaxy can vary dramatically
depending on its position relative to the cluster center.  This is not
simply a reflection of the established morphology-density relationship for
galaxies \citep{dre80}, but rather is evidence that spiral galaxies suffer
from {\HI} stripping near cluster centers.  A similar effect was seen by
\citet{dic97} in the Hercules cluster, where extended {\HI} disks are almost
completely absent in galaxies in some regions but prevalent in others. {\HI}
stripping is seen to be a very efficient process in high density regions such as
rich clusters \citep{cay94}. Consequently, while we will not directly
investigate the efficiency of {\HI} stripping here, the importance of surveying
lower density environments is clear if we are to improve our odds of detecting
galaxies with massive {\HI} disks.   

Our criterion for building a candidate list of potential {\HI}-rich galaxies was
to select APS Catalog galaxies on the blue edge of POSS I color-magnitude
parameter space with apparent magnitude, $m_{E}$, between 16.0 and 21.0.   We
restricted our search to the portion of the sky within 2{\arcdeg} of the PPS
ridgeline defined by \citet{cab98} and within the declination range of the
Arecibo telescope beam.  This is a region west of the A426 and A262 with no
obvious rich clusters.  After removal of plate flaws and scratches identified by
their spurious $O-E$ colors and $E$ mean surface brightnesses, we selected the
30 bluest APS catalog galaxies in each 0.5 $E$ magnitude-wide bin between $E$
magnitude of 16.0 and 20.0.  Using this criterion, we select the bluest 1\% (or
less) of the galaxies in each magnitude bin. From this initial list of 240
candidates on five POSS I fields (P293 through P297, see Table \ref{tblflds}) we
quickly performed a visual inspection of 125 of these candidates chosen to be
the bluest galaxies at each $E$ magnitude. The visual inspection of the POSS I
prints allowed us to reject potential star--galaxy and star--star blended images
(which are prevalent at low Galactic latitudes).  Subsequent examination of POSS
I image parameters of the ``rejected'' candidates showed they had higher surface
brightnesses (lower mean $\mu_E$ values), and had smaller second moments, just
as one would expect for stars.  However, the ``rejected'' candidates were also
typically fainter and smaller, most likely a result of the difficulty of
classifying objects smaller than 0.1 mm in size on the plates by eye.  Using
this method, in one afternoon we reduced the subset of 125 candidates to 71
``clean'' candidates for this {\HI} search, hereafter referred to simply as the
blue edge candidates (see Table \ref{tbl1})   Several candidates are listed
in the NASA/IPAC Extragalactic Database (NED); those cross-identifications are
noted in Table \ref{tbl1}.\footnote{The NASA/IPAC Extragalactic Database (NED)
is operated by the Jet Propulsion Laboratory, California Institute of
Technology, under contract with the National Aeronautics and Space
Administration.}

The PPS lies within 30{\arcdeg} of the Galactic plane, so Galactic extinction is
significant.  Extinction corrections ranged from 0.15 to 0.35 magnitudes in the
O bandpass and 0.09 to 0.19 magnitudes in the E bandpass (see Table \ref{tbl1}).
Thus, we are probing an absolute magnitude range of roughly $-18$ to $-13.5$ in
O, at the PPS distance, fainter than the measured $M_*$ of $-19.4$ for the PPS
\citep{cab98}. No {\it{k}}-corrections have been applied to any of the data
presented in this paper as all the galaxies are below a redshift of 10,000
{\kms} and any such corrections would be less than 0.1 magnitudes.

\section{Arecibo Observations \label{observations}}

We observed 31 of our blue edge candidates, as well as NGC 634 (for
comparison to previous studies), with the 305m Arecibo telescope of the National
Astronomy and Ionosphere Center over 14 nights between August 6 and August 20,
1998 in conjunction with another observing project \citep{cab99}.\footnote{The
National Astronomy and Ionosphere Center is operated by Cornell University under
a cooperative agreement with the National Science Foundation.}  These observed
blue edge candidates were chosen from the full list of 71 blue edge
candidates (listed in Table \ref{tbl1}), based only on their being accessible to
the Arecibo beam during the observing run.  The new Gregorian feed was used with
the narrow L band receiver, resulting in a beam that is roughly oval in shape
(depending on the zenith angle of the observation) with a FWHM of approximately
3{{\arcmin}}.

\subsection{Observational Methods}

We obtained 21cm line spectra centered on the PPS mean velocity of 5000
{\kms} using staggered 25 MHz bands overlapping by 5 MHz, so that our final
velocity coverage was 0 to 10,000 {\kms} (1375 to 1420 MHz). The telescope and
receivers performed very well, with a gain between 6 and 9 K Jy$^{-1}$ and
system temperature between 30 and 35 K depending on zenith angle. The spectra
were calibrated using a previously measured gain function with zenith angle 
(provided by Phil Perillat).

Observations were taken in two on and off pairs, with 4 minutes of integration
on the galaxy position followed by an equal integration on a reference position
offset by 5 minutes in RA.   The initial $\frac{on - off}{off}$ spectra for each
target (resulting from 8 minutes on and 8 minutes off) was stored for
later processing. Some candidates were reobserved and the multiple observations
were co-added to reduce rms noise or to confirm a signal in the presence of
strong radio frequency interference (RFI). In one case a galaxy was detected in
the reference position, so we reobserved the target galaxy with the reference
position taken earlier rather than later (negative offset in RA).

After total telescope time of about 16 minutes per galaxy (for single
observations) we obtained spectra with rms noise of 18 mK in 24 kHz channels,
giving a $3\sigma$ detection threshold of about $5 \times 10^{7} M_{\odot}$ of
{\HI} per channel at 67 Mpc distance assuming the lowest gain of 6 K Jy$^{-1}$.
Fourteen of the 31 blue edge candidates observed were detected in {\HI} and
4 were tentatively detected.  These detections are detailed in Section
\ref{rogues}.

\subsection{Managing Radio Frequency Interference}

Radio Frequency Interference (RFI) is the bane of extragalactic spectroscopy in
the centimeter-wave band, particularly for single dish surveys like this one.
Strong RFI can completely compromise an entire scan, but much more common is
weak interference that is confined to a fairly narrow frequency range. These
weak RFI signals can and do appear throughout our band.  Some are fairly
constant, and are thus easily distinguishable, while others are intermittent.
Some appear and disappear in the on$-$source and off$-$source positions just as
a real galaxy would.  Some even reappear at roughly the same time on different
days, so they can seem to be repeatable from day to day as a real galaxy must
be.  Thus it is not possible to definitively distinguish between a real
extragalactic {\HI} emission line and an interfering terrestrial signal by any
means.  

However, a very useful diagnostic for separating RFI from galaxy emission lines
is the difference between the two circular polarizations.  Since terrestrial
interference is typically somewhat linearly polarized, any difference between
the signal in the two polarizations identifies it as likely interference
({\it{e.g.}} - Schneider 1996). Our method for masking out interfering signals is
based on an interactive task which displays the eight spectra (four bands times
two polarizations) for each scan together, and then allows flagging of
individual channels or channel ranges in one or both polarizations.  

In addition to generating spurious detections, RFI consistently covered certain
frequency ranges, which means that our non-detections must be qualified by
details of which velocity ranges were not searchable due to consistent RFI.  As
an overall guide, Figure \ref{cumilrfi} shows the aggregate of all frequencies
which were blanked due to interference in our entire run.  These were determined
for each source observation separately.  Most spectra have three to five
interference signals between 1375 and 1405 MHz. Figure \ref{cumilrfi} shows that
the frequency range 1388.4 to 1388.6 MHz (6865 to 6910 {\kms}) is hardly ever
usable, while the range from 1380 to 1382 (8300 to 8800 {\kms}) is often covered
as well.   These velocity ranges are therefore effectively excluded from our
search.

In a few cases, it was difficult to determine whether we had observed weak RFI
or a narrow emission line from a galaxy. In such cases we have tried to
reobserve on different days to confirm the detection, but this was not always
feasible.  We class such observations as tentative detections (Group 3 in Table
\ref{tbl2}). It is also possible that interference can blend with a real galaxy
detection to cause errors in the line integral, center velocity, and velocity
width measurements.  We indicate any such suspicious values for these measurements
in Table \ref{tbl2}.

\subsection{Review of Detected Galaxies \label{rogues}}

Of the 31 blue edge candidates we observed, 19 (or 61\%) are detected (or
tentatively detected) in {\HI}.  In addition to these 19 detections, we detected
an additional three galaxies in off-scans and obtained a serendipitous detection
of UGC 630 in the field of MAPS-P295-1915216.\footnote{The IAU designation of
objects in the Minnesota Automated Plate Scanner Catalog indicates the POSS I
field number, in this case P295, and the unique O plate ``starnum'' identifying
that object in that field, in this case 1915216.}

The detected galaxies were divided into five groups to aid in the initial
analysis.  Group 1 consists of {\HI} detections with high signal-to-noise ratios
and contains 14 of our {\HI} galaxies and the serendipitously observed galaxy,
UGC 630 (see Figure \ref{group1}). Group 2 consists of positions around
MAPS-P295-1369071, which were repeatedly observed in order to confirm that our
detection is not confused by another galaxy on the edge of the field with the
Arecibo beam (see Figure \ref{group2}). Group 3 contains the four tentative
{\HI} detections (see Figure \ref{group3}). Group 4 consists of the three
galaxies we detected in the off$-$scans of our galaxies (see Figure
\ref{group4}). And group 5 contains NGC 634, which was observed for comparison
with previous observations. We discuss the detected galaxies individually below.

The 3{\arcmin} Arecibo beam does not resolve the galaxies and thus provides no
direct information on the rotation curves of these {\HI} disks.  Our goal has
been detection, not high-resolution {\HI} imaging. However, many dwarf galaxies
which have been mapped with aperture synthesis telescopes show rising rotation
curves, not only throughout their optical disks, but throughout their {\HI}
extent as well \citep{tay96}. In the extreme case of a solid body rotation curve
throughout the {\HI} disk, the resulting single dish spectrum should have a
semi-circular or semi-elliptical shape, with a single peak at the center
velocity.  If the rotation curve flattens at large radii this shape changes to
the familiar "two-horned" profile, with two peaks at the two velocity extrema of
the line, which correspond to the projected rotation velocities of the two ends
of the major axis (e.g. Lavezzi and Dickey 1997). Most of our observed galaxies
have profiles which have a square-top or two-horned shape, suggesting that their
rotation curves are flattening at large radii.  This needs to be confirmed with
aperture synthesis mapping.

Our individual observations are detailed below, in the order in which they
appear in Table \ref{tbl2}.

{\it MAPS-P293-268610}: This galaxy was observed twice.  The first time its
spectrum was confused by emission from the galaxy UGC 105 in the reference
position.  We reobserved it the following day, with the reference spectrum taken
at $\alpha-5^{m}$ instead of $\alpha+5^{m}$, to sort out the two profiles.  In
fact there is no overlap in velocity between the two galaxies, so we have
averaged the two observations here, blanking the channels below 8250 {\kms},
which are covered by UGC 105. The profile appears to have two peaks (``horns'')
with sharp edges on either side from which we may infer that this is a rotating
gas disk with a rotation curve that flattens at large radii.  This inference
assumes a ``typical'' {\HI} surface density profile for a disk galaxy and will
require confirmation via aperture synthesis observations.

{\it MAPS-P293-100987}: This galaxy, also identified as F409-05 from
\citet{ede89}, shows a two-horned profile shape, indicating an extended, flat
rotation curve in the outer regions (assuming an {\HI} distribution as seen in
disk galaxies).  While our velocity, 4851 {\kms}, is consistent with the
velocity of 4864 {\kms} reported by \citet{ede89}, our gas mass,
$1.8\times10^{9} M_{\odot}$, is a bit lower than the value of $2.26\times10^{9}
M_{\odot}$ we compute from Eder {\etal}'s data.

{\it MAPS-P293-249758}: This galaxy was observed twice on different days, and
the results are consistent, however on both days there appears to be a narrow
interference spike near the center of the line profile.  The presence of this
interference in the middle of such a faint line raises the concern that the
entire profile may be interference generated, either by sidebands of the signal
or by the spectrometer resolving function.  However that interference signal is
present throughout the entire night of observing on both days, and it never
shows width of more than about four channels (1388.5 to 1388.6 MHz).  So
although our measurements of this profile suffer from uncertainty due to the
necessity of masking these channels in the middle of the presumed galactic
line, we are moderately confident of the detection.

{\it MAPS-P293-179271}  This clearly detected line comes from a very faint
galaxy, whose optical image is almost blended with a pair of bright stars.  The
line profile is single peaked, which suggests a disk with a solid body rotation
curve, a face-on orientation, or perhaps a turbulent cloud of gas. 

{\it MAPS-P294-727319}: This faint, diffuse galaxy has an extremely bright {\HI}
line, which indicates a gas mass of roughly $2\times10^{9} M_{\odot}$.  Yet the
line is centrally peaked, which implies that it comes from a cloud or disk which
does not have a flat rotation curve, but may be more or less in solid body
rotation. A warped disk is one possible explanation of the line shape, although
confirmation will, again, require aperture synthesis observations.

{\it MAPS-P295-699736}: This is a strong and very well detected line.  We
repeated the observation on two different days with consistent results, and
there is no interference nearby in frequency on either day.  Optically the
galaxy is extended, but everywhere low in surface brightness, just above the
plate limit in both colors.  This is a good example of a massive gas disk with
very low optical luminosity.

{\it MAPS-P293-1790367}: This galaxy was observed twice, with consistent results
between the two days.  There is a suggestion of two ``horns'', but these may
correspond to two systems which appear separate on the optical image.

{\it MAPS-P295-823940}: This galaxy was observed on two different days, with
consistent results. However the second day's observations were compromised by
interference on the low velocity side of the galaxy (5250 to 5350 {\kms}). 
Such a narrow, single-peaked profile could easily be generated by an interfering
signal.  We have moderate confidence that this does represent a detection, but
there is a significant chance (perhaps 10 percent) that we are being fooled by
an interference signal.

{\it MAPS-P295-1693362}: This relatively bright galaxy (cross-identified in NED
as CGCG 501-047) shows a very clear two-horned profile shape.  The gas disk
probably has an extended, flat rotation curve in the outer parts.  The optical
image has a bright nucleus and an extended disk.

{\it MAPS-P295-1915216}: The spectrum in this direction shows two emission
features which look like galaxies.  The stronger one (at 4878 {\kms}) has a
classic two-horned shape, indicating a disk in circular rotation with a rotation
curve which flattens at the outer edges.  The second feature (at 6786 {\kms}) is
weaker, but still very well detected above the noise.  The optical image shows a
relatively large galaxy (identified in NED as UGC 630) about an arc minute to
the west of the blue dwarf which was our target.  UGC 630 has a measured
redshift of 4884 {\kms} \citep{sch90}, we therefore identify the other (6786
{\kms}) line with MAPS-P295-1915216.

{\it MAPS-P295-642385}: This galaxy was observed on two different days, with
consistent results.  On one day there was interference at higher velocities
(4850 to 5000 {\kms}), but the second day is very clean in this area, so we are
very confident of the detection.  The line is very narrow and single-peaked,
which suggests solid body rotation throughout the HI, if there is a disk in
circular rotation.  Several faint, diffuse images appear to the south west of
the galaxy, all within the beam area, so the correspondence of the line with the
central object is in some question.

{\it MAPS-P295-1202937}: This is a very strong detection of an extremely faint
galaxy. The line profile is clearly two-peaked, corresponding to an extended
disk of gas.  We have not had time to map around this position, but there seem
to be no alternative galaxies in the vicinity which could account for the
observed gas.

{\it MAPS-P295-1326244}: This is a very narrow line from a relatively bright
galaxy.  The detection is somewhat tentative as we have not had time to confirm
it with a reobservation. The narrow width makes it somewhat more likely than
most of our detections to be interference.  Assuming it is real, the shape of
the profile suggests either a disk seen nearly face on, or one with a
solid-body rotation curve.

{\it MAPS-P295-910484}:  This strong line is a clear-cut detection of this
diffuse, blue galaxy. The sharp edges on the line, and the suggestion of two
peaks, imply an extended disk of gas with a rotation curve which flattens at
the edges.

{\it MAPS-P295-1369071}: This strong line was observed only on one day, but it is
far from any interference and the detection is quite solid.  The line shape
suggests a disk with an extended, flat rotation curve. The mass of gas in this
galaxy is so high, and its optical luminosity so low, that we have searched the
vicinity of the object to confirm that the gas cloud is centered on the optical
candidate.  We observed three other positions, one on either side of this object
offset by 1.3{\arcmin}, and one centered on the nearby galaxy NPM1G $+30.0027$.
The results confirm that this faint blue galaxy is the center of the HI cloud.
The spectra on either side have similar profiles, but weaker by about a factor
of 0.6, as expected given the beam shape.  The emission toward NPM1G $+30.0027$
is much fainter yet.  Thus the identification of this line with the faint blue
galaxy MAPS-P295-1369071 appears to be conclusive.  The extended Arecibo
observations also allow us to determine that the direction of the galaxy's spin
vector is northward. The relationship among these beam centers and their
corresponding spectra are illustrated on Figure \ref{group2}.

{\it MAPS-P295-2259922}: This galaxy exhibits is a very weak line.  It appears
at about the five sigma level in two different observations taken on two
different days.  The results of the two are consistent within the errors.  On
neither day is there obvious interference at this frequency, although on the
first day there is interference at somewhat higher velocity (6000 {\kms}).  We
have confidence that this is a real emission line, but given its low
signal-to-noise we class it as a tentative detection.

{\it MAPS-P295-83632}: This is a very tentative detection, the most questionable
of all of our ``lines.''  The profile resembles a weak interference feature. 
Although the signal-to-noise ratio is good, we are very hesitant to identify
this feature as real HI emission because of its narrow width and very sharp
edge.  We have not had time to follow-up with a reobservation on a second day.

{\it MAPS-P295-773634}: This weak line is a very tentative detection.  It was
observed only once, we did not have time to return for another observation on
another day.  The line appears identically in both polarizations, and there is
no interference nearby.  The baseline is quite flat, so it is unlikely that
this could be an artifact of the receiver or spectrometer. However the line is
near the radiometer noise limit, so its parameters cannot be well measured.

{\it MAPS-P295-576122}: This weak line is a very tentative detection.  It was
observed only once.  There is a narrow interference feature about 150 {\kms}
higher, but it seems to be confined to an extremely narrow band.  We have not
had time to reobserve this position to confirm the detection.

{\it UGC 105}: Detected in the reference beam for MAPS-P293-268610.  We found
{\HI} associated with UGC 105, 78{\arcsec} from the reference beam center.  Its
identification is supported by the match of our velocity to the velocity of 8046
{\kms} from the Third Reference Catalogue of Bright Galaxies (\citet{deV91},
hereafter RC3) catalog \citep{deV91}.

{\it MAPS-P293-102675}: A serendipitous detection in the reference beam, a
relatively bright galaxy 50{\arcsec} from the reference position of
MAPS-P293-100987 appears to have been detected, as there are no other likely
candidates in the field of the reference beam.

{\it MAPS-P294-444433}:  An apparent serendipitous detection at the reference
position of MAPS-P295-1577104, this curious feature appeared as a pair of
negative lines in our observation of this galaxy.  It is in a fairly
interference-free region of the band, although there is a narrow spike about
100 {\kms} higher in velocity than the center of the higher velocity emission
line.  The nearest galaxy to the reference position of the observation appears
to be MAPS-P294-444433, located 83{\arcsec} from the beam center. The detection
is only tentative, and is not obviously associated with the galaxy, but it is
worthy of follow-up observations. If it is real, it may represent either two
objects, or a disk or ring of gas.

{\it NGC 634}: We observed this galaxy as a test of the system.  It has been
described before by \citet{weg93} and by \citet{the98}.  The central velocities
in the literature are 4942 and 4925 {\kms}, we find 4916 {\kms}, which is in
good agreement, given the broad {\HI} line width.  \citet{weg93} find widths
of 520 {\kms} at 50\% of the mean and 492 {\kms} at 20\% of the peak versus our
measurement of 512 {\kms}.  We measure 4.5 Jy-{\kms} for the line integral;
\citet{the98} find 4.3 Jy-{\kms} but \citet{weg93} find 6.8 Jy-{\kms}.  Thus our
calibration seems to be in rough agreement with earlier work, although the range
of values for the line integral is a bit worrisome.

\section{{\HI} Properties of the blue edge Galaxies \label{BEgas}}

The {\HI} masses of our (detected) blue edge galaxies range from
$2.7\times10^8 M\solar$ to $3.6\times10^9 M\solar$.\footnote{We do not attempt
to derive the dynamical masses for our blue edge galaxies because the
resolution of the POSS I plates for images this small results in large
uncertainties in the galaxy's inclination angle.} These gas masses fall about 1
dex below the turnoff in the {\HI} mass function \citep{sch96}, but on the high
end of the gas masses typically seen for dwarf irregulars such as the SMC
\citep{ski96}.  This is consistent with their integrated {\HI} spectral
profiles, which generally show two-horned profiles typical of galaxies with flat
rotation curves.  Such flat rotation curves are common for larger spirals and
LSBs (\citet{ver01} and references therein) as opposed to dwarf galaxies, whose
rotation curves typically rise throughout their disks \citep{tay96}.

In order to estimate the gas mass-to-light ratios of our blue edge galaxies,
we compute the distance modulus as
\begin{equation}
m-M = 25 - 5 \log H_0 + 5 \log cz  + 1.086(1-q_0)z
\label{distmod}
\end{equation}
\citep{wei72}, where we assume $H_0$ = 72 {\kms} and $q_0$ = 0.5.  No
corrections have been made for Virgo infall or internal extinction.  Using this
definition of the distance modulus, we find that our blue edge galaxies have
gas mass-to-light ratios, $M_{\HI}/L_{O}$, of 0.09 to 4.79 in solar units.  If
$M_{\HI}/L_{O} \approx M_{\HI}/L_{B}$, this indicates most of our blue edge
galaxies have gas mass-to-light ratios considerably higher than typical HSB
spiral galaxies, which have $M_{\HI}/L_{B}$ typically below 0.60 (in solar
units) \citep{rob94, mai98}.  Such high gas mass-to-light ratios are typical of
LSBs \citep{imp96,one97a}.  This suggests our blue edge galaxies may have
{\HI} properties common to LSBs, something we will investigate more throughly in
Section \ref{distinct}.

Before continuing, we briefly address an issue we mentioned in Section
\ref{probing}: the possibility of {\HI} stripping in galaxies near the PPS
ridgeline. The {\HI} mass deficiency of galaxies is known to correlate with
environmental density \citep{gio85}.  Although we do not have density estimates
for the PPS per se, we previously defined a high surface density ``ridgeline''
based on the distribution of the CfA Redshift Catalog (hereafter ZCAT,
\citet{huc92}, 2000 February version) galaxies in the PPS field \citep{cab98}.
Because the PPS is oriented edge-on \citep{gio88}, angular distances from the
ridgeline correspond to distances from the highest density portions of the PPS.
Our investigation finds no correlation between {\HI} mass and angular distance
from the PPS ridgeline (see Figure \ref{ridgeHI}a).  A very weak correlation
($r=0.26$) between the {\HI} mass to light ratio, $M_{\HI}/L_{O}$, and angular
distance from the ridgeline is observed (Figure \ref{ridgeHI}b). The best-fit
line describing this correlation is 
\begin{equation}
M_{\HI}/L_{O}(r_d) = 0.613(\pm 0.564) \times r_d + 1.750(\pm 0.061), \label{bestfit}
\end{equation}
where $r_d$ is the ridgeline distance in degrees.  However, the best-fit slope
is only 1.1$\sigma$ from a value of zero, there is a significant likelihood
there is no correlation at all.

\section{Are the blue edge Galaxies a Distinct Population? \label{distinct}}

This project started with the realization that the majority of the optical
counterparts to the {\HI}-selected galaxies from \citet{dic97} lay on the ``blue
edge'' of the POSS I color-magnitude diagram for galaxies.  Our {\HI}
observations show the converse is also true: many blue edge galaxies appear
to be {\HI}-rich.  This leads us to ask in what other ways these ``blue
edge'' galaxies are distinct from the ``normal'' HSB galaxy population (other than
being bluer)?  The answer, as we will show, is that the 19 blue edge
galaxies we detected in {\HI} (those in Groups 1, 2, or 3) are clearly a distinct
population from ``normal'' HSB galaxies in both their optical and {\HI}
properties.  We support our claim using several previously published HSB and LSB
galaxy catalogs (Table \ref{CompCats}) to construct four major pieces of
evidence that we will outline in this Section.  First, the bivariate brightness
distributions of our blue edge galaxies are more similar to those of LSBs
than HSBs. Second, known LSBs appear as blue edge galaxies on the POSS I. 
Third, the {\HI} masses of our blue edge galaxies are much lower than
typical HSB galaxies, but typical of known LSBs at comparable redshifts.  And
finally, the gas mass to light ratio of our blue edge galaxies are similar
to known LSBs of similar POSS I color. 

\subsection{The Bivariate Brightness Distribution of blue edge Galaxies \label{BBDs}}

\citet{dri00} suggest an approach to classifying galaxies based on their
positions in a bivariate brightness distribution (hereafter BBD) plot, a plot of
mean surface brightness versus absolute magnitude.  They use this approach to
show that the traditional Hubble sequence of galaxies segregates reasonably well
into different regions of this parameter space.  Furthermore, non-Hubble
sequence galaxies such as LSBs and cDs are easily accommodated on the BBD plot
and appear isolated from Spirals and Ellipticals.  Thus, a BBD plot provides a
simple means of comparing the optical properties of the blue edge galaxies
we observed with those of previously identified HSB and LSB galaxies.  With this
goal in mind, we cross-identified several galaxy catalogs containing redshifts
with the APS catalog (those listed under the ``BBD'' column of Table
\ref{CompCats}). Correcting the $O$ magnitudes (but not diameters) for Galactic
extinction as outlined in Section \ref{HercReview}, we estimated the absolute
$O$ magnitude (using equation \ref{distmod}) and mean $O$ surface brightness for
each comparison galaxy.

A BBD of our blue edge galaxies and a large sample of ``normal'' HSB
galaxies in the PPS field extracted from the ZCAT clearly shows that the two populations differ (See Figure
\ref{ourbivar}). We can account for the fact that our blue edge galaxies are
intrinsically fainter than the ZCAT by noting that while not explicitly
flux-limited, the ZCAT contains mostly bright galaxies with $O<17$. However, our
blue edge galaxies are also lower surface brightness than the ZCAT galaxies.
This is intriguing since surface brightness was not a criterion for identifying
blue edge galaxies. By adding POSS I data for LSB galaxies identified from
another plate-based survey \citep{imp96} and a CCD-based survey (\citet{one97a},
hereafter OBC) to our previous BBD plot, we find that our blue edge galaxies
lie on the high luminosity end of the LSB population in the BBD plot. This
implies blue edge galaxies are very similar galaxies to the LSBs seen on the
POSS I (Figure \ref{bivariate}).

\subsection{Known LSBs in POSS I Color-Magnitude Parameter Space \label{LSBsOnBE}}

Driven by the fact that the BBD of our blue edge galaxies overlaps with that of
known LSBs, we now examine the POSS I color-magnitude distributions of
previously identified LSBs, including several from LSB catalogs with no
redshifts available (see the ``color-magnitude'' column of Table
\ref{CompCats}).  A plot of the color-magnitude distribution of these LSBs shows
that the majority of them also lie along the blue edge (see Figure
\ref{ALLcolormag}).  We quantify this by defining the color difference from the
blue edge boundary as
\begin{equation}
\Delta(O-E)_{BE} = (O-E) - (O-E)_{BE}
\label{BEdiff}
\end{equation}
where $(O-E)$ is the extinction corrected $O-E$ color of the galaxy and the
$(O-E)_{BE}$ is defined by equation \ref{BEeqn}.  Using equation \ref{BEdiff},
we determine the cumulative $\Delta(O-E)_{BE}$ distributions for each of these
LSB galaxy catalogs (Figure \ref{BEdistribution}).  All these catalogs show
$\Delta(O-E)_{BE}$ distributions very different from the general POSS I galaxy
population, but quite similar to each other.  Whereas only 10\% of APS catalog
galaxies have $\Delta(O-E)_{BE}<0$ (by definition), over 68\% of LSBs galaxies
lie blueward of the blue edge boundary!  Surprisingly, this is even the case
for the LSBs from the OBC catalog, which includes many LSBs with red $B-V$
colors.  This apparent discrepancy will be addressed in Section \ref{BEexplain}.

The maximum difference of the cumulative $\Delta(O-E)_{BE}$ distributions
between datasets can be used as a Kolmogorov--Smirnov (K--S) D statistic to
compute the likelihood that the datasets were drawn from the same parent
distribution (see \citet{pre92}).  Such K--S tests indicate that these LSBs have
probabilities of less than $10^{-15}$ of being drawn from the same
$\Delta(O-E)_{BE}$ distribution as the POSS I galaxies (See Table \ref{probBE}).
 This doesn't mean that all these LSBs have identical $\Delta(O-E)_{BE}$
distributions (a quick examination of Table \ref{probBE} shows they don't), but
they are clearly all drawn from populations that lie preferentially on the
blue edge of POSS I color-magnitude parameter space.

\subsection{Comparison of {\HI} Properties with Known LSBs \label{gascomp}}

We now compare the {\HI} properties of our blue edge galaxies (as outlined
in Section \ref{BEgas}) with those of three other galaxy catalogs: a high
surface brightness catalog of bright galaxies (the RC3) and two LSB catalogs
(see the ``$M_{\HI}$'' column of Table \ref{CompCats}). Since a given radio flux
limit will result in a varying minimum {\HI} mass limit with varying redshift,
we restrict the comparison galaxies to the observed range of redshifts of our
blue edge galaxies (4000 {\kms} $<cz<9000$ {\kms}).  To ensure that our
comparison samples are not biased toward fainter galaxies, we also require that
they exhibit the same optical flux limit of $m_E<20$ as the blue edge
galaxies. Applying these restrictions results in relatively small comparison
samples of LSBs (see Table \ref{CompCats}), but ensures they have similar
optical flux and {\HI} mass limits. 

Our blue edge galaxies exhibit $M_{\HI}$ values peaked around $\log(M_{\HI})
\sim 9.0$.  This peak lies roughly 0.8 dex below the peak observed for HSB
galaxies from the RC3, but lies in the range of the $M_{\HI}$ values seen
for LSBs (Figure \ref{HIdistros}a). We quantify the level of resemblance between
these distributions of $M_{\HI}$ values by using K--S tests to compare them (see
Table \ref{probM}).  As expected based on Figure \ref{HIdistros}, our ``blue
edge'' galaxies have an $M_{\HI}$ distribution that is most similar to the LSBs,
with the OBC galaxies having an 88\% chance of being drawn from the same parent
$M_{\HI}$ distribution. This stands in marked contrast to the HSB galaxies in
the RC3, which have only a probability of $3\times10^{-10}$ of sharing the
same parent $M_{\HI}$ distribution as the blue edge galaxies.  Clearly,
based on these small samples, our blue edge galaxies share more in common
with LSBs than HSBs where there gas masses are concerned.

In retrospect, this should not be a very surprising result. If galaxies exhibit
a limited range of gas mass-to-light ratios, we expect that the more luminous
galaxies will tend to have higher total gas masses.  This expectation is
confirmed in a plot of $M_{\HI}$ versus $L_O$ for these galaxies (Figure
\ref{lumHI};Table \ref{probM2L}).  The RC3 galaxies all lie at relatively high
luminosities and have correspondingly high $M_{\HI}$ values.   Our blue edge
galaxies lie at the low luminosity, low gas mass edge of the RC3 distribution.
Their distribution does not extend to the very lowest luminosities seen for LSB
galaxies. Notably, the distribution of $M_{\HI}$ versus $L_{O}$ for these
galaxies has a slope less than one, indicating higher gas mass-to-light ratios
for less luminous galaxies, an observation previously made by \citet{sta92}. A
separate examination of the $M_{\HI}/L_O$ distribution of our blue edge
galaxies reveals it is peaked around $\log(M_{\HI}/L_{O}) \sim 0.25$ (Figure
\ref{HIdistros}b). This is significantly higher than the peak $M_{\HI}/L_O$ for
the RC3 galaxies ($\sim0.5$ dex higher) but appears to be roughly equal to the
peak of the distributions for the LSBs.  This is supported by K--S tests
comparing the distributions which show both LSB catalogs have a 20\% or greater
chance of sharing the same gas mass-to-light distribution as our blue edge
galaxies, while the HSB galaxies from the RC3 show only a 0.3\% chance.   On the
other hand, it appears that our blue edge galaxies lack the extremely high
$M_{\HI}/L_O$ galaxies seen in LSB catalogs (as seen in Figure \ref{lumHI}),
though this conclusion is based on a population of only 19 blue edge
galaxies.  

We might be able to explain this bias against extremely high $M_{\HI}/L_O$ blue
edge galaxies by considering that color and $M_{\HI}/L_O$ may be correlated.
\citet{bot84} found that the $M_{\HI}/L_{B}$ of late-type spirals was correlated
with $B-V$ and even more strongly with $B-H$.  Bothun ascribed these
correlations to a relationship between the initial star formation rate (SFR) of
the galaxy and its subsequent evolution \citep{bot82}.  The better correlation
using infrared bandpasses was attributed to the fact that blue luminosities of
galaxies are more affected by uncertain levels of internal extinction. 
\citet{gir87} instead argues that the correlation using infrared bandpasses is
better because $H$ bandpass luminosity depends on an older population of stars
than $B$ bandpass luminosity and thus  $M_{\HI}/L_{H}$ is a measure of the
present {\HI} reservoir of the galaxy versus its star formation history (rather
than its current SFR).  More recently, \citet{mat97} find that for extreme
late-type galaxies, the blue galaxies all have high $M_{\HI}/L_{V}$, while red
galaxies exhibit a broad range of gas mass-to-light ratios. Matthews and
Gallagher suggest that this broad range in $M_{\HI}/L_{V}$ for red galaxies is
because some galaxies with older stellar populations retain large {\HI}
reservoirs in their outer disks, where surface densities drop below the Toomre
criterion for star formation to occur.

Using the four catalogs cited above (see the $M_{\HI}$ column of Table
\ref{CompCats}), we construct a plot of $M_{\HI}/L_{O}$ versus $O-E$ (see Figure
\ref{colorM2Lall}). This plot reveals a trend similar to that cited by
\citet{mat97} in that redder galaxies tend to have a wider distribution of
$M_{\HI}/L_{O}$ than blue galaxies.   From this plot, it is clear our ``blue
edge'' galaxies mark the bluest end of a flared distribution of $M_{\HI}/L_{O}$
versus $O-E$.   From this plot, we see that the LSBs reaching gas mass-to-light
ratios much higher than our blue edge galaxies are all relatively red. And
from examining this diagram, we see our blue edge galaxies lie in a region
dominated by very blue LSBs, which have a very similar range of smaller
$M_{\HI}/L_{O}$ ratios.   In a model proposed by \citet{mat97}, this suggests our
galaxies have a lower likelihood of having large untapped {\HI} reservoirs in
their outer disks than the redder LSBs.

\subsection{The blue edge Galaxies as LSBs}

Classifying a galaxy as a {\it{bona fide}} LSB traditionally requires deep
images to assess its surface brightness profile to see if it meets specified
``low surface brightness'' criterion.  Without such images for our blue edge
galaxies, we can not absolutely verify their LSB nature.  However, we assert that our
blue edge galaxies as a population seem very similar to LSBs in most
respects.  Most previously identified LSBs visible on the POSS I lie on the
blue edge of POSS I color-magnitude parameter space.   The bivariate
brightness distribution of blue edge galaxies appears relatively similar to
that of known LSBs on the POSS I.  Furthermore, the {\HI} properties of our
blue edge galaxies exhibit similar $M_{\HI}$ distributions to LSBs at
similar redshifts. If we take into account the relationship between
$M_{\HI}/L_{O}$ and $O-E$ (Figure \ref{colorM2Lall}), it is clear that our
blue edge galaxies exhibit very similar gas mass-to-light ratios
as blue LSBs. Thus, despite the fact that low surface brightness was not an
explicit selection criterion for identifying the blue edge galaxies, the
evidence at hand suggests that our blue edge galaxies share much in common
with the LSBs visible on the POSS I.  Of course, the relatively high surface
brightness limit of the POSS I means LSBs visible in that survey are likely to
represent the high surface brightness end of the LSB population. Thus, we would
be safer to identify our blue edge galaxies as similar to the high surface
brightness end of the LSBs.

\section{Understanding the blue edge \label{BEexplain}}

The fact that LSBs lie on the blue edge of color-magnitude parameter space
is not surprising, since LSBs are observed to be very blue \citep{bot97,imp97}.
Thus, given our selection criterion for blue edge galaxies it is plausible
that we are selecting blue {\HI}-rich LSBs. However, the preponderance of blue
LSBs has been questioned by O'Neil and collaborators who recently performed a
CCD-based search for LSBs \citep{one97a} and discovered several LSB galaxies
with red $B-V$ colors.  \citet{one97b} suggest that the observed preponderance
of blue LSBs is a selection effect due to the blue sensitivity of photographic
plates used in most previous optical LSB searches.  Therefore it was surprising
that our cross-identification of the 73 OBC galaxies on the POSS I fields which
were online in the APS Catalog in August 2000 recovered 58 of them. 
Furthermore, these OBC galaxies are predominantly on the blue edge of the
POSS I color-magnitude space (as seen in Figure \ref{BEdistribution}).  Are we
truly seeing LSBs with ``red'' $B-V$ on the ``blue edge?''

\subsection{Red LSBs on the blue edge?}

If we assume, as \citet{one97b} do, that the blue sensitivity of photographic
plates is a strong selection effect, then a cross-identification of the OBC with
the POSS I would preferentially match only the bluest galaxies in the OBC
survey.  However, a plot of $B-V$ vs. $V$ for OBC galaxies separated into
matched and unmatched galaxies shows the APS Catalog does not preferentially
match the bluest OBC galaxies (see Figure \ref{distrocomp}a).  While the average
$B-V$ color of the unmatched galaxies is slightly redder (${\langle}B-V{\rangle}
= 0.78\pm0.01$) than that of matched galaxies (${\langle}B-V{\rangle} =
0.82\pm0.02$), the $B-V$ color distribution of the matched galaxies extends both
blueward and redward of that of the unmatched galaxies.  Instead, it appears the
dominant bias in the POSS I is that it doesn't recover the lowest surface
brightness OBC galaxies (Figure \ref{distrocomp}b).   If we can understand how
OBC galaxies with ``red'' $B-V$ colors can appear on the blue edge of POSS I
color-magnitude parameter space, we may have an explanation for why LSBs appear
on the ``blue edge.''

A plot of $O-E$ vs. $B-V$ shows that these two colors appear to be weakly
correlated (with a correlation coefficient of 0.22, see Figure \ref{colorcomp}).
In fact, the OBC galaxy with the reddest $B-V$ has the second most blue $O-E$
color! This discrepancy between $O-E$ and $B-V$ colors for OBC galaxies is
especially startling when compared to the behavior of galaxies in the RC3, which
show a significant correlation between $O-E$ and $B-V$ colors. For 618 RC3
galaxies with $B$ and $V$ color information cross-identified with the APS
catalog (see the ``photometry'' column of Table \ref{CompCats}), we find a
correlation coefficient of 0.68 between APS $O-E$ and RC3 $B-V$ colors. The
radically different photometric behavior of the LSBs from OBC and the HSB
galaxies in the RC3 can be understood if we consider the blue and red bandpass
behavior separately.

We expect that the blue $O$ and $B$ magnitudes of galaxies should be correlated
because the $O$ bandpass covers the entire $B$ bandpass and extends a bit
blueward. This expected correlation is observed for the 617 higher surface
brightness RC3 galaxies with both $B$ and $O$ measurements available (and has a
correlation coefficient of $r=0.90$). In contrast to the overlapping blue
bandpasses, the $E$ bandpass lies about 1300{\AA} redward of $V$, centered near
the {\Ha} line at 6562{\AA}.  Consequently, any correlation between the $V$ and
$E$ magnitudes might be expected to be weaker. However, we find that $E$ and $V$
magnitudes for the RC3 are only slightly less well correlated ($r=0.88$) than
the $O$ and $B$ bandpasses.  What is surprising is that similar investigation of
the OBC galaxies shows much weaker correlations between $O$ and $B$ ($r=0.66$)
and $E$ and $V$ ($r=0.61$). A direct comparison of the two samples shows that if
we fit the relationship for RC3 galaxies $B$ versus $O$ magnitudes with a linear
fit, the fainter and more diffuse OBC galaxies lie at systematically fainter
POSS I $O$ magnitudes than this fit would predict (See Figures
\ref{magcomparison}a). The faintest OBC galaxies appear to have $O$ magnitudes 1
to 2 magnitudes fainter than suggested by their $B$ magnitudes if we extrapolate
the RC3 relationship between these two magnitudes.  The situation with the $E$
magnitudes is more extreme, with the roughly half of OBC galaxies appearing to
be 2 to 3 magnitudes fainter in $E$ than predicted based an extrapolation of the
RC3 relationship between $E$ and $V$ \ref{magcomparison}b). It is the difference
in these two effects that results in the OBC galaxies having $O-E$ colors
considerably bluer than their $B-V$ colors would suggest (Figure
\ref{magcomparison}c).

\subsection{An Explanation for the blue edge}

Since the whole field of LSBs was opened up by Disney's realization that surface
brightness was being generally ignored as a selection effect \citep{dis76}, let
us consider the possibility that the blue edge effect is due to the surface
brightness limits of the POSS I plates. The APS Catalog is constructed from
digital scans of the POSS I performed in threshold densitometry mode where data
is kept only for regions with photographic densities greater than 65\% the sky
density, resulting in a threshold surface brightness limit for objects in the
APS catalog slightly deeper than the sky surface brightness on each plate (See
Table \ref{tblflds}). It has been established by the APS Project that the
limiting surface brightness, $\mu(lim)$, of the POSS I $O$ (blue) plates is
typically deeper than that of the $E$ (red) plates.   This is not an issue when
comparing high surface brightness galaxy photometry where the majority of the
galaxy's luminosity comes from HSB regions.  However, it is possible for a LSB
galaxy with neutral color and mean surface brightness near the plate limit to
barely appear on the $E$ plate while clearly appearing on the $O$ plate due to
the deeper surface brightness limits.  This will result in a reported bluer
$O-E$ than an otherwise similar HSB galaxy.\footnote{In fact, this represents a
previously known selection effect, since extremely blue low surface brightness
objects will not appear on the $E$ plate at all, and thus will not reside in the
APS catalog.} This suggests that LSBs may be biased to lie on the blue edge
of POSS I color-magnitude parameter space simply as a result of the fact that
$\mu_{E}(lim) < \mu_{O}(lim)$ and not necessarily because they are intrinsically
blue.

To support this theory, we investigated the
differences between the $O$ and $B$ (and $E$ and $V$) magnitudes for the OBC
galaxies versus $\mu_{B}(0)$, their reported $B$ central surface brightness (see
Figures \ref{magVmu}a and \ref{magVmu}b).  In Figure \ref{magVmu}a we see that
the difference between the $O$ and $B$ magnitudes is mildly affected by the
$\mu_{B}(0)$ of the OBC galaxy, with a computed best-fit line of the form
\begin{equation}
O-B = -7.874(\pm 3.901) + 0.383(\pm 0.171)\mu_{B}(0). \label{O-BvMU}
\end{equation}
Figure \ref{magVmu}b shows the corresponding relationship for $E-V$ is
considerably stronger with
\begin{equation}
E-V = -21.281(\pm 6.079) + 0.984(\pm 0.266)\mu_{B}(0). \label{E-VvMU}
\end{equation}
The fact that the relationship between these magnitude differences and central
surface brightness is steeper for the redder bandpasses ($E$ and $V$) is simply
a reflection of the fact that $\mu_{E}(lim) < \mu_{O}(lim)$.   The net result is
that the $O-E$ colors of LSBs are systematically bluer on the POSS I than they
would be on a survey achieving equal surface brightness limits in both red and
blue bandpasses.

In passing, we note that the POSS I blue plates (with their deeper surface
brightness limits) appear to have a very linear relationship between $B$ and $O$
up to $B\sim18.5$ (Figure \ref{magcomparison}a), meaning our estimates of the gas
mass-to-light ratios of our blue edge galaxies should not be strongly
affected by this problem.  However, any future attempts to probe for LSBs close
to the plate limit may have to compensate for this bias in the $O$ magnitudes of
low surface brightness objects near the plate limit.

\subsection{Red LSBs on the POSS I}

Given this systematic bias for LSBs to appear blue on the POSS I, what can we
say about known LSBs that are observed to be red in $O-E$? Returning to Figure
\ref{ALLcolormag} we notice that there is a substantial population of LSBs that
have colors significantly redward of the blue edge boundary (defined in
equation \ref{BEeqn}). An investigation of these very red LSBs shows they in
truth are red.  First of all, while in almost all cases the mean $O$ (blue)
surface brightnesses of these galaxies are lower than their mean $E$ (red)
surface brightnesses, red galaxies exhibit larger differences between the blue
and red surface brightnesses just as one would expect (see Figure
\ref{colorVparam}a).  Furthermore, if the galaxies are truly red, then we would
expect their $O$ plate images to be smaller than their $E$ plate images. 
Despite the fact that the $E$ plates usually have shallower surface brightness
limits than the $O$ plates, we confirm that the $O$ plate images for the red
LSBs are smaller than the $E$ plate images (see Figure \ref{colorVparam}b). 
These two facts together imply that these red galaxies are indeed red LSBs. 

\section{Conclusions}

We have developed a simple criterion for identifying {\HI}-rich galaxy
candidates from the POSS I based on their positions in POSS I bivariate
color-magnitude parameter space.  While the method described here relies on
followup visual inspection of candidates to eliminate star--galaxy and
star--star blended images, subsequent investigation of the image parameters of
the ``rejected'' shows they consist mainly of images with higher surface
brightnesses  and smaller second moments, suggesting automation of our selection
method is possible.  Such automation should allow us to identify fainter and
smaller blue edge candidates than is practical with visual inspection.  We
observed 31 of our 71 blue edge candidates and detected 19 (61\%) in {\HI}
(15 with high signal-to-noise).

The blue edge galaxies we detect in {\HI} are very distinct from ``normal''
HSB galaxies in both their optical and {\HI} properties.   We have concluded
that our blue edge galaxies are very similar to LSBs in these properties,
based on cross-identification of known LSBs with the POSS I. The evidence
supporting this tentative identification of blue edge galaxies as LSBs
includes several key points: 
\begin{enumerate} 
\item Cross-identification of known LSBs with the POSS I shows
that over 68\% of them lie blueward of the blue edge boundary (defined in
equation \ref{BEeqn}). While this certainly doesn't mean blue edge galaxies
need be LSBs, it suggests that many LSBs lie on the blue edge of POSS I
color-magnitude parameter space.
\item The bivariate brightness distributions of known LSBs and our blue edge
galaxies are similar, with both populations lying at lower mean surface
brightness than HSBs drawn from the ZCAT (Figure \ref{bivariate}).
\item The {\HI} masses of blue edge galaxies are typically lower and their
gas mass-to-light ratios higher than those of luminous HSB galaxies (drawn from
the RC3).  We find that their gas masses are very similar to those of LSBs at
the same redshift, although their gas-mass-to-light ratios do not extend to the
highest values seen for LSBs.  If we take into account the relationship between
$O-E$ color and $M_{\HI}/L_{O}$ seen in Figure \ref{colorM2Lall}, we see that
blue galaxies appear to have a more limited range of gas mass-to-light ratios
than redder galaxies, and blue LSBs exhibit a similar range of $M_{\HI}/L_{O}$
as our blue edge galaxies.
\item The two-horned profiles seen for many of our blue edge galaxies are
typical of galaxies with flat rotation curves.  Such flat rotation curves are
generally associated with large spirals and LSBs \citep{ver01}.  This combined
with their typical {\HI} mass of $\log(M_{\HI}) \sim 9.4$ suggests these are not
dwarf galaxies, but rather disk systems.
\end{enumerate} 
Based on these pieces of evidence, we believe these blue edge galaxies are
likely to be LSBs on the high surface brightness end of that population (so as
to be visible on the POSS I).  They are likely to be disk systems and not the
smaller dwarf galaxies commonly associated with LSBs.  This suggests our ``blue
edge'' criteria selects the same type of galaxies a blind radio survey
\citep{dic97} or wide-field optical survey \citep{imp97,one97a} for {\HI}-rich
LSBs would select.

While it has been noted that LSBs are among the bluest (non-starbursting)
galaxies known \citep{imp97}, this alone does not appear to be the explanation
for why LSBs lie preferentially on the blue edge of the POSS I
color-magnitude diagram. Specifically, it cannot explain why the ``red'' (in
$B-V$) LSBs from \citet{one97a} can be blue in $O-E$ (Figure
\ref{BEdistribution}).  We demonstrate that this is most likely the result of
differing surface brightness limits on the POSS I $O$ (blue) versus $E$ (red)
plates, which make low surface brightness galaxies appear systematically bluer.
This does not mean LSBs are not truly ``blue,'' it simply means that when
working near the limiting surface brightness of a survey, the integrated fluxes
must be measured to equal isophotes in all bandpasses in order to assure the
colors are not systematically biased.  This bias for low surface brightness
objects to appear blue on the POSS I means that the blue edge is an
excellent filter for sifting out LSBs and similar objects from the large number
of galaxies in the field.

This model for why  LSBs lie on the blue edge of the POSS I leads to an
important question" just what regions of these LSBs are detected on the POSS I
plates? Answering this question will require detailed comparison of POSS I
images with deep CCD images.  The POSS I most likely only detects the highest
surface brightness regions of any LSB.  This leads to another interpretation of
our blue edge galaxies: they could be in a class with NGC 2915, a galaxy
which has been identified as a blue compact dwarf but has the {\HI} properties
of a ``dark'' spiral galaxy \citep{meu96}. Only observations of blue compact
dwarfs deeper than the POSS I plate limits can reveal if they have extremely low
surface brightness disks surrounding the detected POSS I images, presumably
current star forming regions in an otherwise extremely low surface brightness
galaxy.

Finally, we see no simple mechanism to make low surface brightness galaxies
appear systematically redder on the POSS I versus CCD observations. We have
shown that known LSBs with red $O-E$ colors tend to appear smaller on the $O$
plate and have lower $O$ mean surface brightnesses than blue LSBs (Figure
\ref{colorVparam}), indicating that these galaxies are indeed red.  Employing
simulations of the ISM of LSBs, \citet{ger99} were able to reproduce the optical
properties of LSBs only when assuming LSBs have a sporadic star formation rate.
de Blok's simulations suggest that roughly 80\% of LSBs have some current star
formation with young, luminous stars dominating the luminosity and leading to
their observed blue colors.  The remaining quiescent LSBs should appear red
($B-V>1$ or approximately $O-E>0.7$) as they would only have an older stellar
population. And while we know the POSS I colors of LSBs are biased to be
systematically blue, an examination of Figure \ref{BEdistribution} shows that
approximately 50\% of all POSS I galaxies have $O-E>0.7$, while only $\sim10\%$
of LSBs on the POSS I do.  These redder LSBs could be currently in a quiescent
stage in their evolution and as such represent interesting candidates for
followup if we want to search for LSBs with old stellar populations as predicted
by \citet{ger99}.

\acknowledgments

We would like to thank Roberta Humphreys and Evan Skillman for discussions of
potential astrophysical causes for the ``blue edge.'' Thanks to Steve Odewahn
for an illuminating conversation on observational systematics in the POSS I and
to the anonymous referee for his or her helpful suggestions.  JEC thanks the
Department of Physics, Astronomy, and Engineering Science at Saint Cloud State
University who provided him with support when some of this research was
conducted and Bruce Partridge for reviewing a late draft of this paper.

We express our gratitude to Karen O'Neil and Jo Ann Eder for providing
electronic versions of the data from their respective low surface brightness
galaxy catalogs.

We would like to thank telescope operators Miguel Boggiano, Willie Portalatin,
Pedro Torres, and Norberto Despiau for their good humor and help with observing
(and especially Norberto for his ``lucky coffee'').  JEC would like to thank
Chris Salter, Tapasi Ghosh, Jo Ann Eder, and Phil Perillat for helping make his
first radio observing experience excellent, both professionally and personally.
Travel was sponsored by the National Astronomy and Ionosphere Center and
the University of Minnesota Graduate School.

This research was financially supported by the University of Minnesota and NSF
grants AST 97-32695 and AST 00-71192. This research has made use of the APS
Catalog of the POSS I, which is supported by the National Aeronautics and Space
Administration and the University of Minnesota. The APS databases can be
accessed at {\it http://aps.umn.edu/} on the World Wide Web.

We acknowledge the use of NASA's {\it SkyView} facility ({\it
http://skyview.gsfc.nasa.gov}) located at NASA Goddard Space Flight Center.

\clearpage



\figcaption[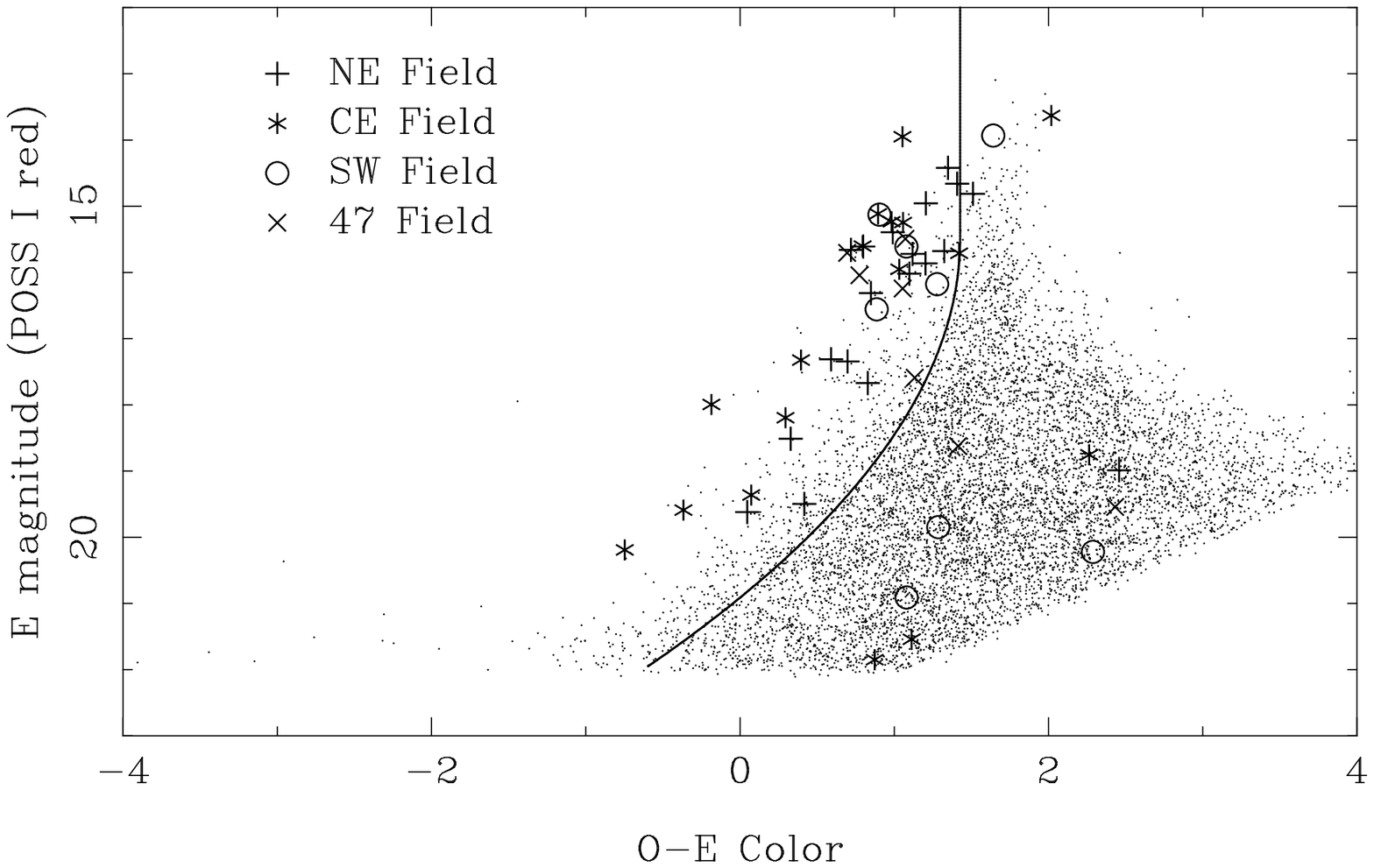]{A POSS I color-magnitude diagram of galaxies in the Hercules
cluster from the four fields in \citet{dic97}.  Notice that the vast majority of these
{\HI}-rich galaxies lie on the blue edge of the color-magnitude diagram.  Note:
An approximate color transformation for $O-E$ to more contemporary bandpasses
is $O-E \approx B-R+0.30$ as noted by \citet{cab98}. \label{colormag}}

\figcaption[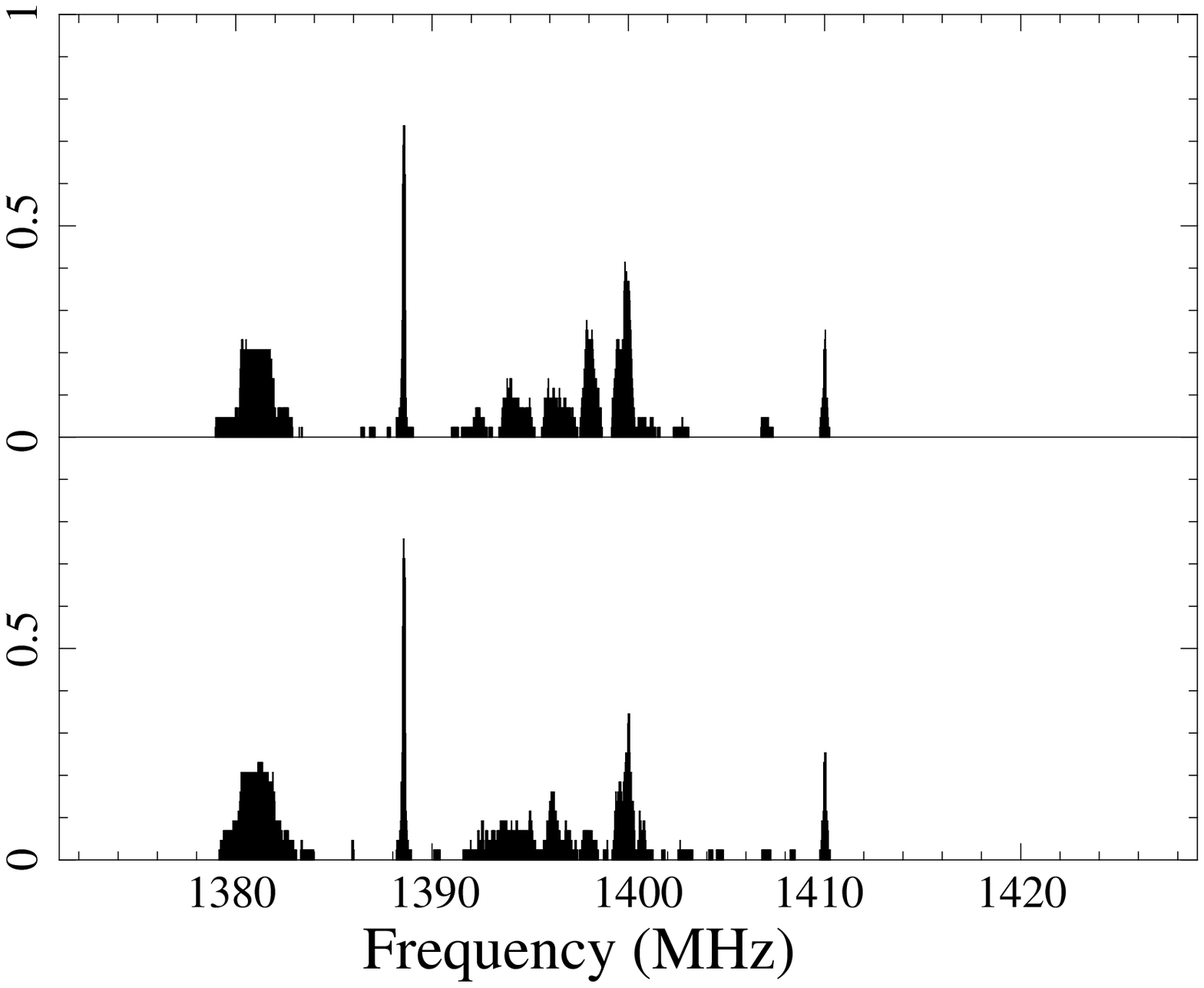]{The final histogram of the total fraction of
observing time which was flagged due to radio frequency interference versus
frequency. The top (bottom) panel shows right (left) circular polarization.
\label{cumilrfi}}

\figcaption[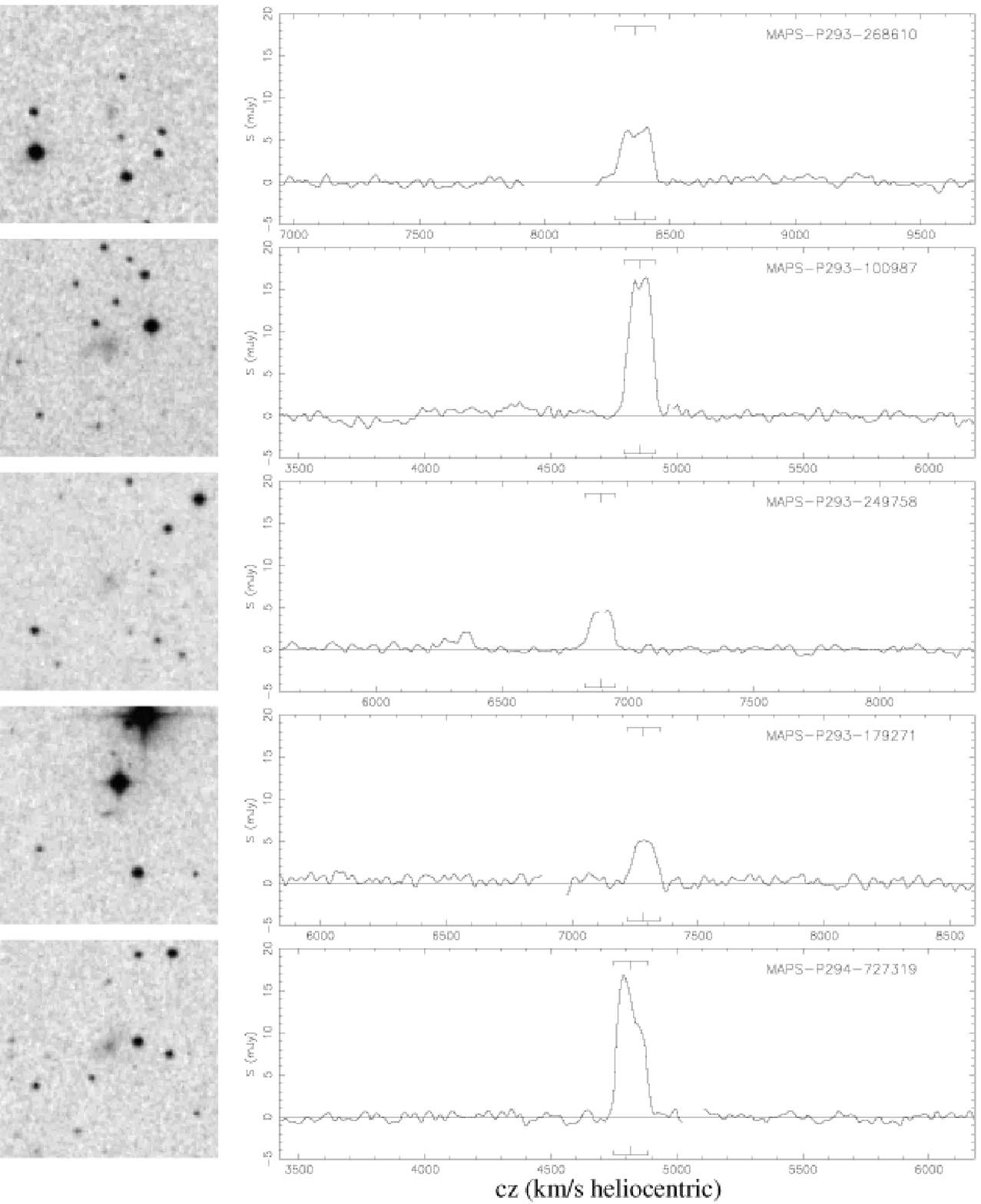]{{\HI} spectra and 3{\arcmin} by 3{\arcmin} POSS I (Red)
images of candidates detected in {\HI} including NGC 634.  Segments of the 
spectral profile determined to be due to interference are blanked. \label{group1}}

\figcaption[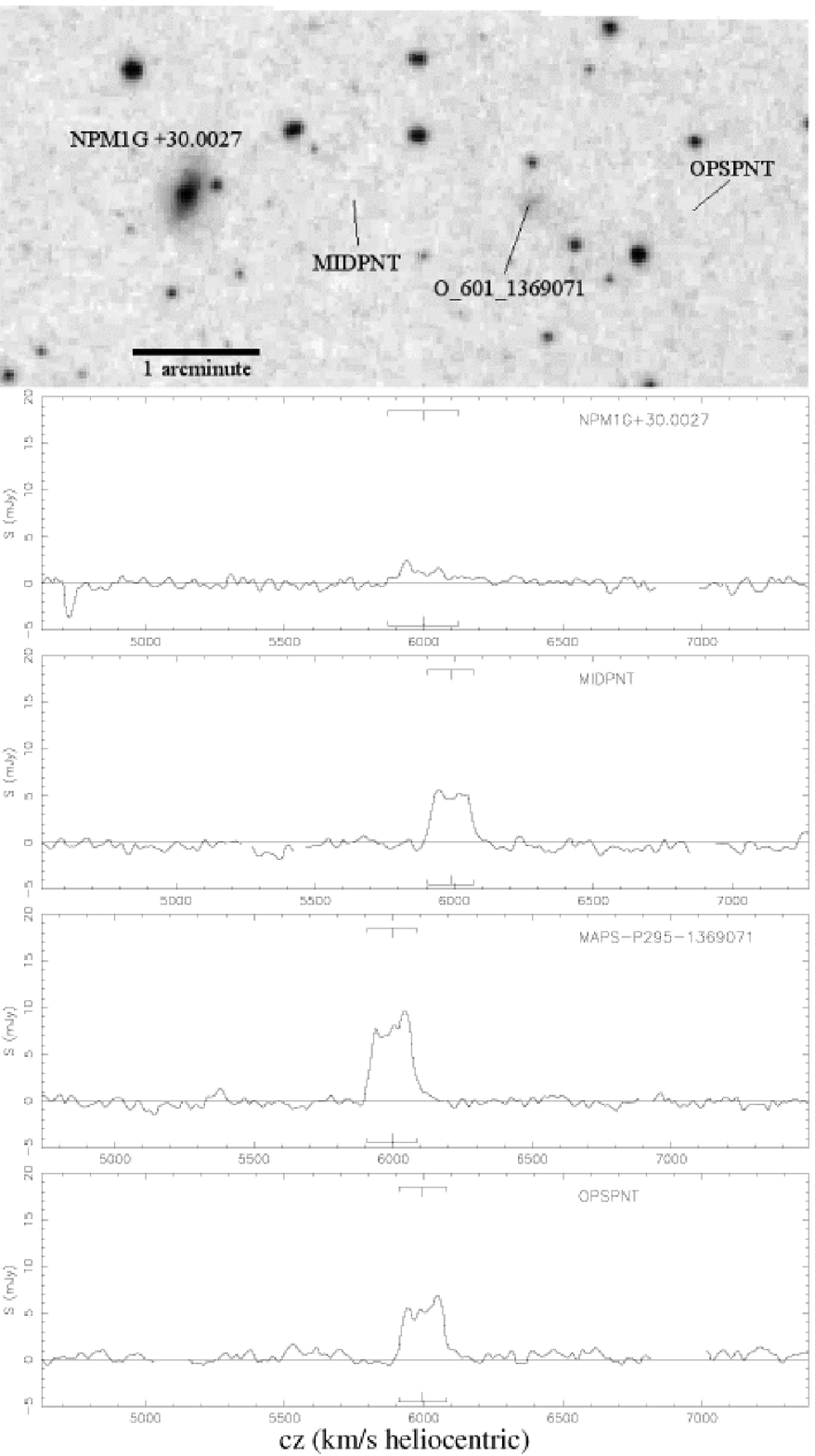]{The four {\HI} spectra and POSS I (Red) image of
the region surrounding MAPS-P295-1369071, where there was the possibility of
confusion with NPM1G +30.0027. As before, segments of the spectral profile
determined to be due to interference are blanked. \label{group2}}

\figcaption[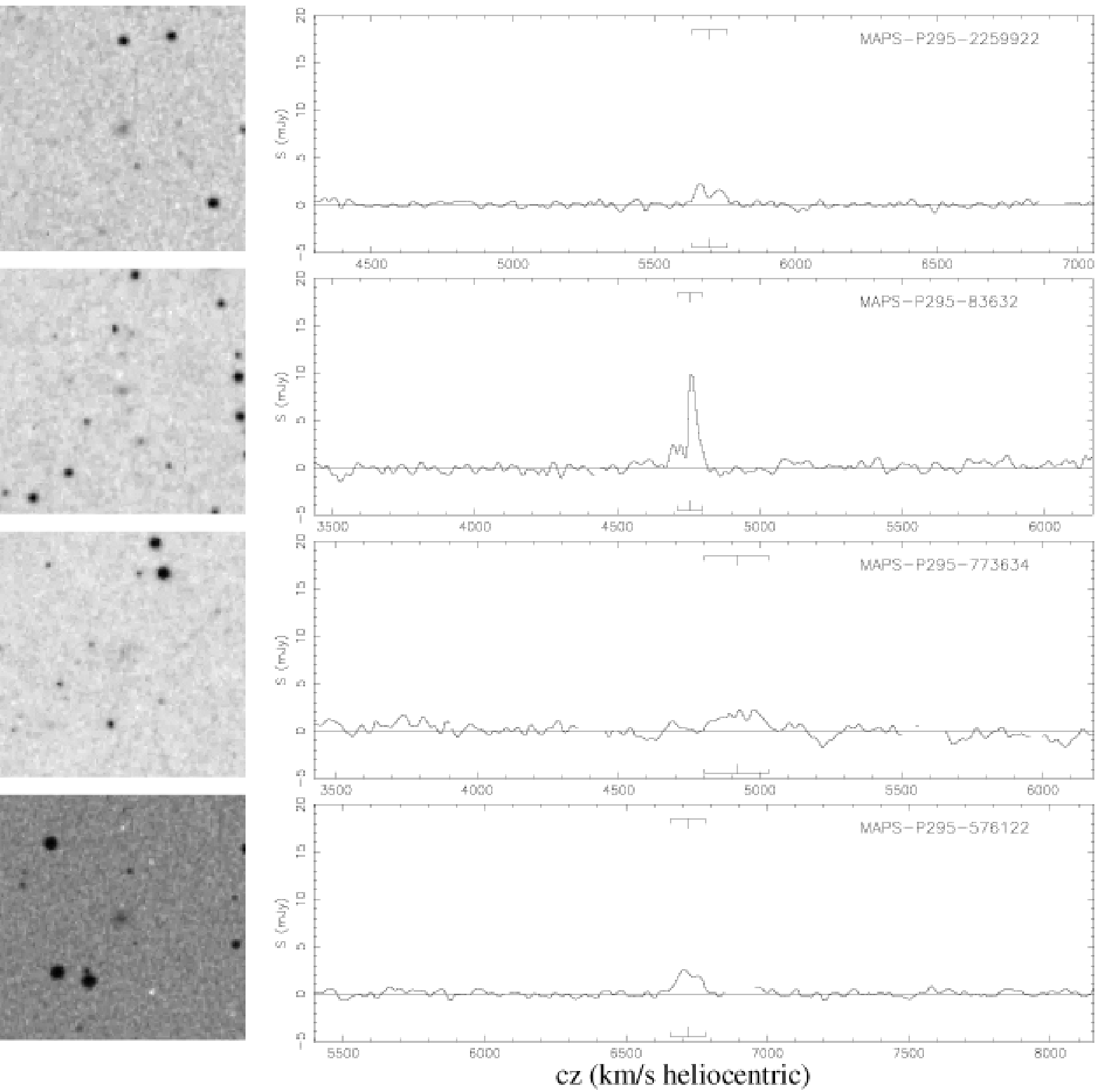]{Like Figure \ref{group1}, but for galaxies
tentatively detected in {\HI}.\label{group3}}

\figcaption[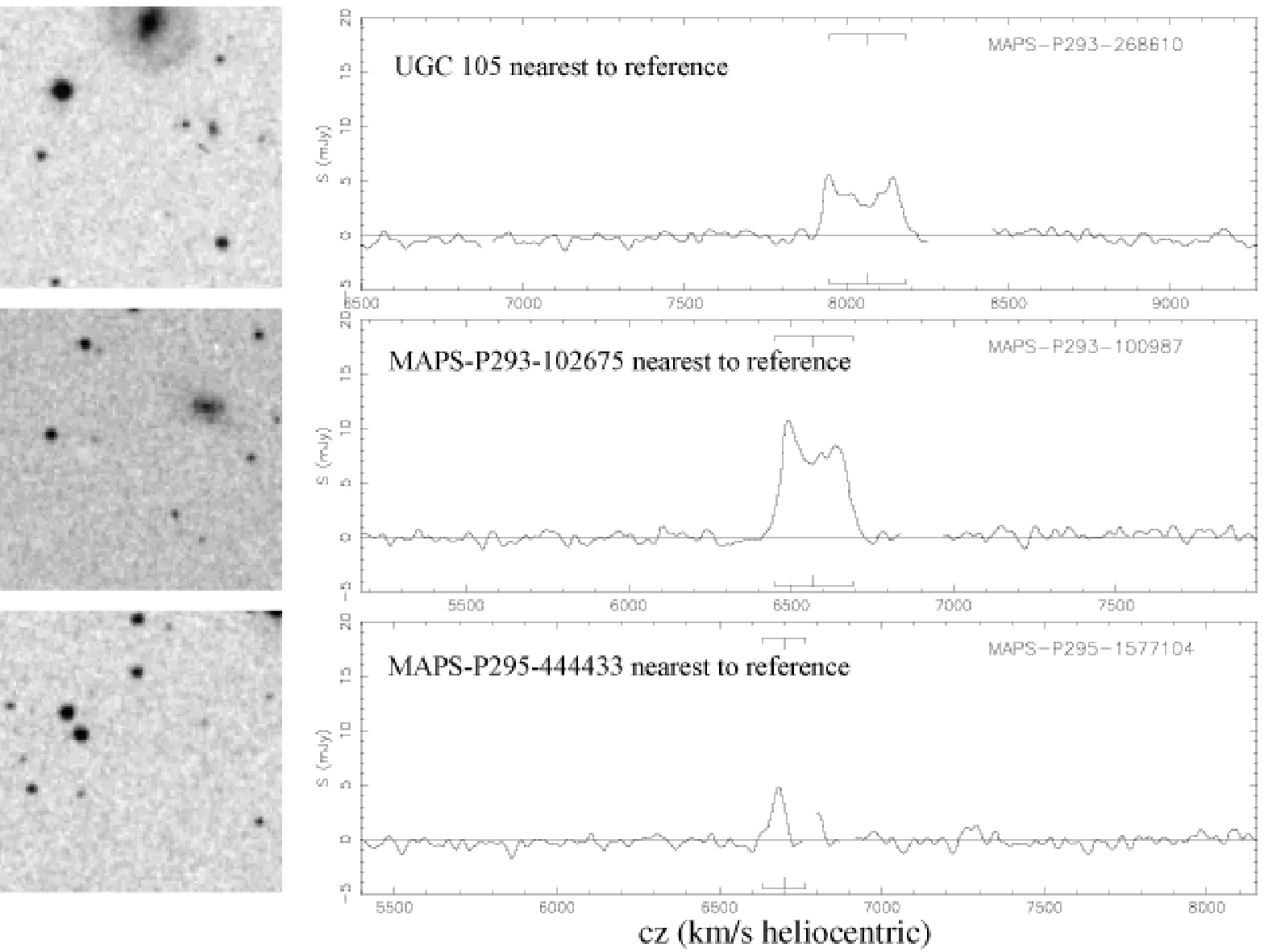]{Like Figure \ref{group1}, but for candidates
detected in {\HI} in the off scans.  Images are of the off scan positions in
each case. \label{group4}}

\figcaption[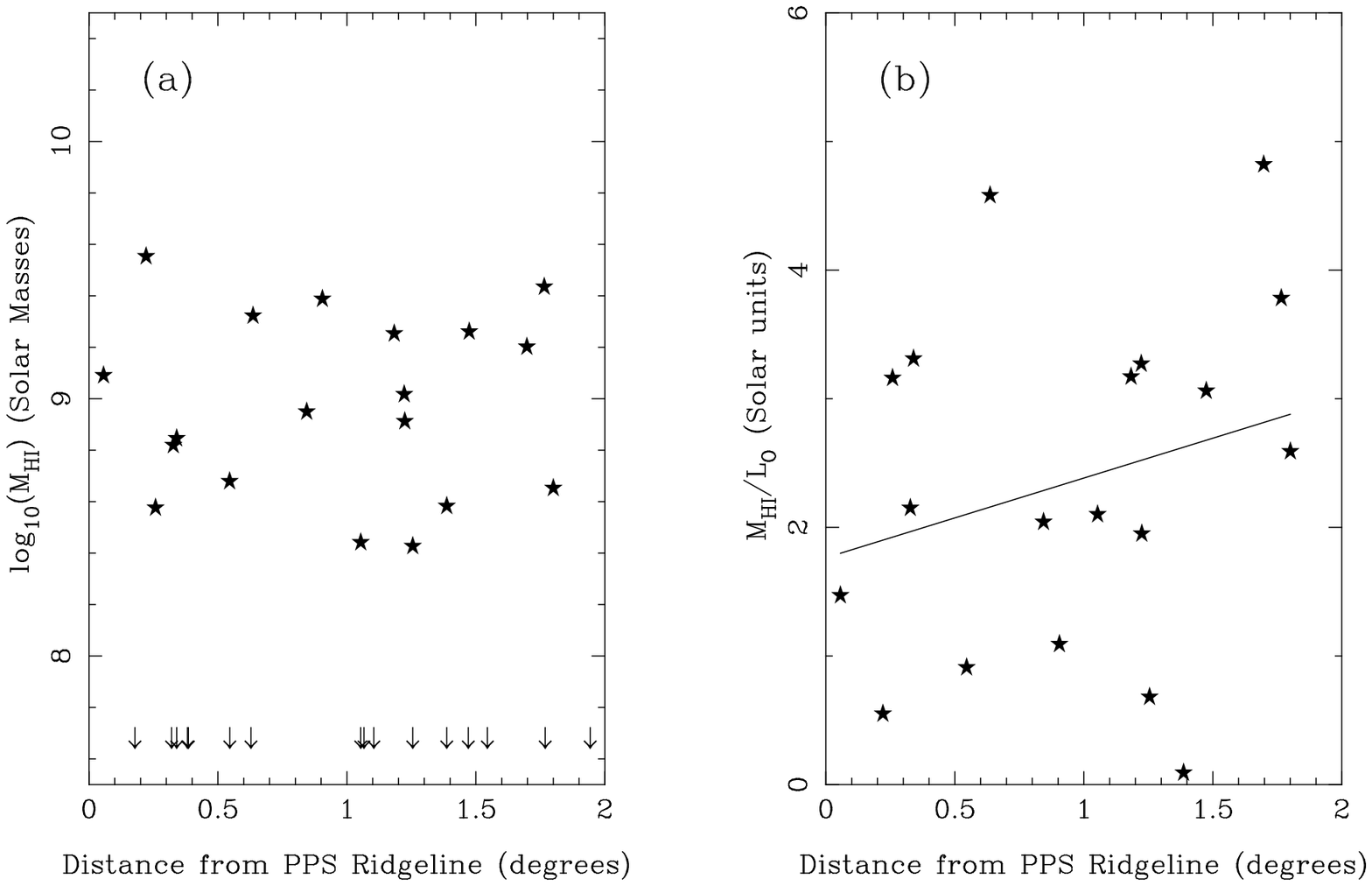]{(a) A plot of the logarithm of the {\HI} mass (in
units of solar mass) versus the angular distance (in degrees) from the PPS
ridgeline \citep{cab98} for the blue edge galaxies. All objects which were
undetected are shown as downward arrows indicating the estimated upper {\HI}
mass limit for the observed blue edge galaxies not detected in {\HI}. (b) A
plot of  $M_{\HI}/L_{O}$ versus the angular distance from the PPS ridgeline for
the blue edge galaxies.  The best fit line is shown and is described by
equation \ref{bestfit}.  This best fit line refects a weak correlation
($r=0.255$) between $M_{\HI}/L_{O}$ and distance from the PPS ridgeline.
\label{ridgeHI}}

\figcaption[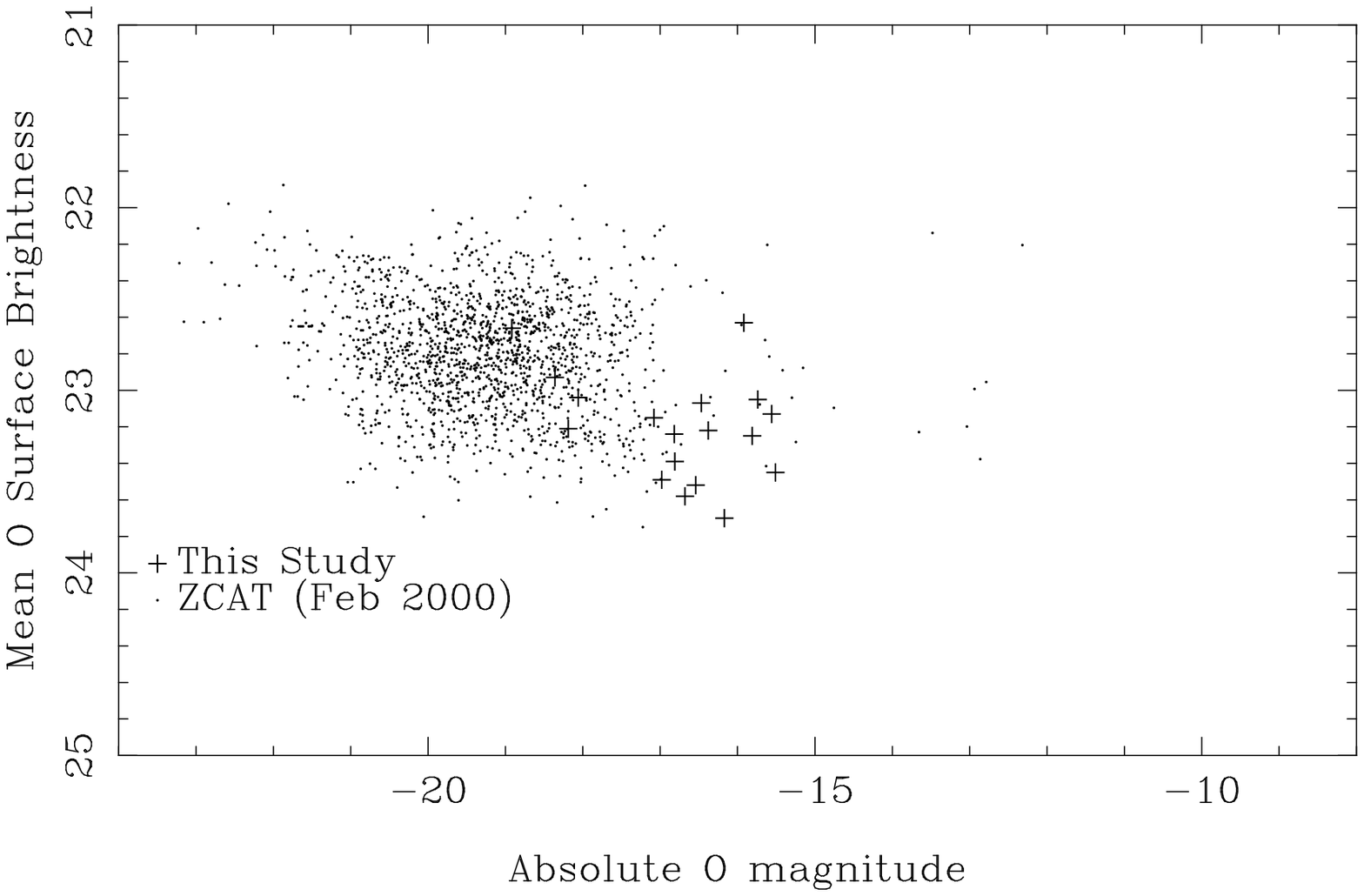]{The ``Bivariate Brightness Distribution''
(BBD) for our observed blue edge galaxies versus the ZCAT.  Notice that the
blue edge population appears to be fainter and of lower surface brightness
than the ZCAT population (labeled with dots).  The separation in luminosity is
expected given the flux-limits of the ZCAT, but distinction in surface
brightness between the two populations hints that our blue edge galaxies are
a distinct galaxy population from those in the ZCAT. \label{ourbivar}}

\figcaption[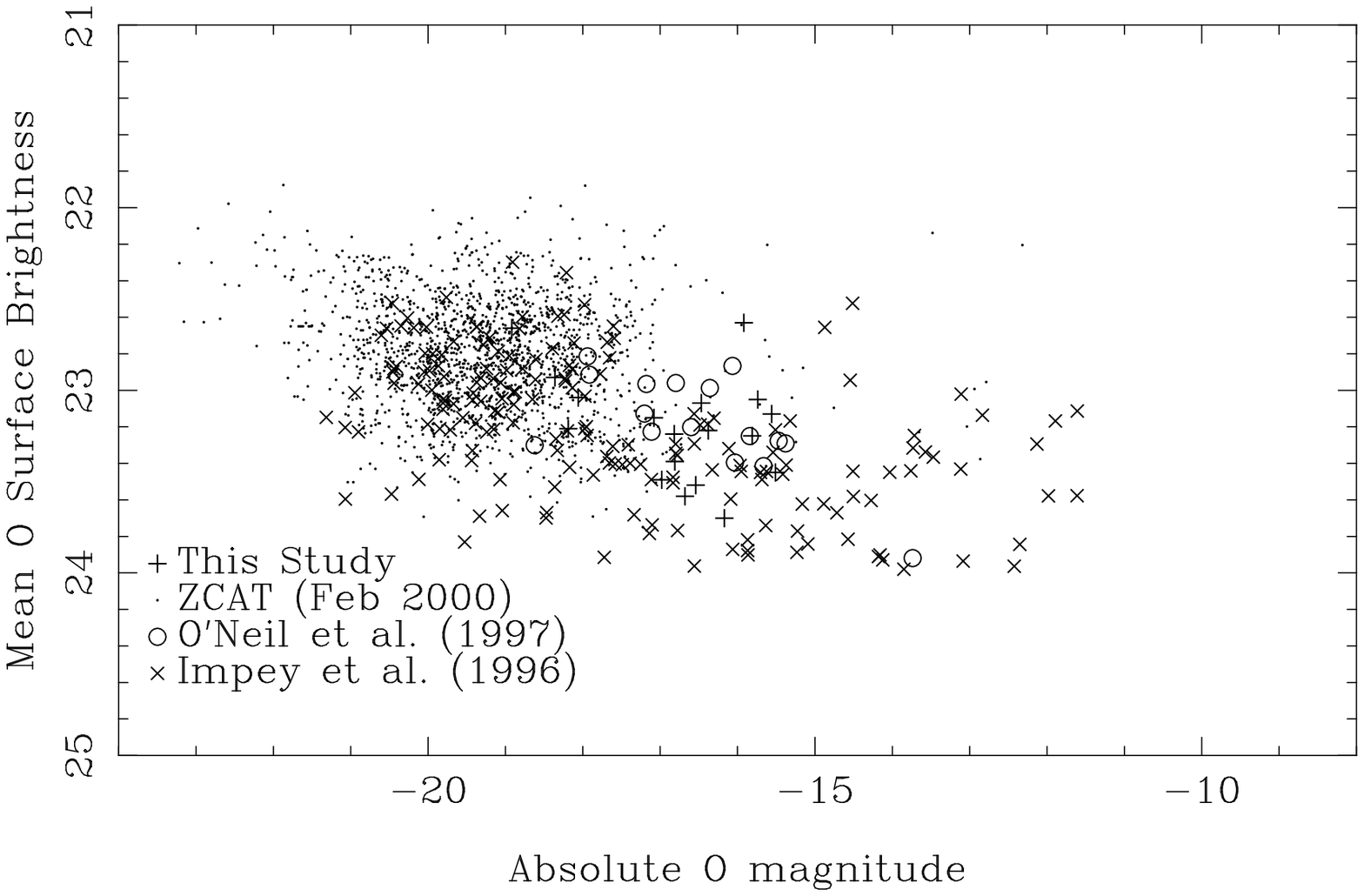]{BBD plot like Figure \ref{ourbivar} but adding the
\citet{one97a} and \citet{imp96} LSB catalogs. Our blue edge galaxies
(labeled with crosses) lie among the LSB catalog galaxies. \label{bivariate}}

\figcaption[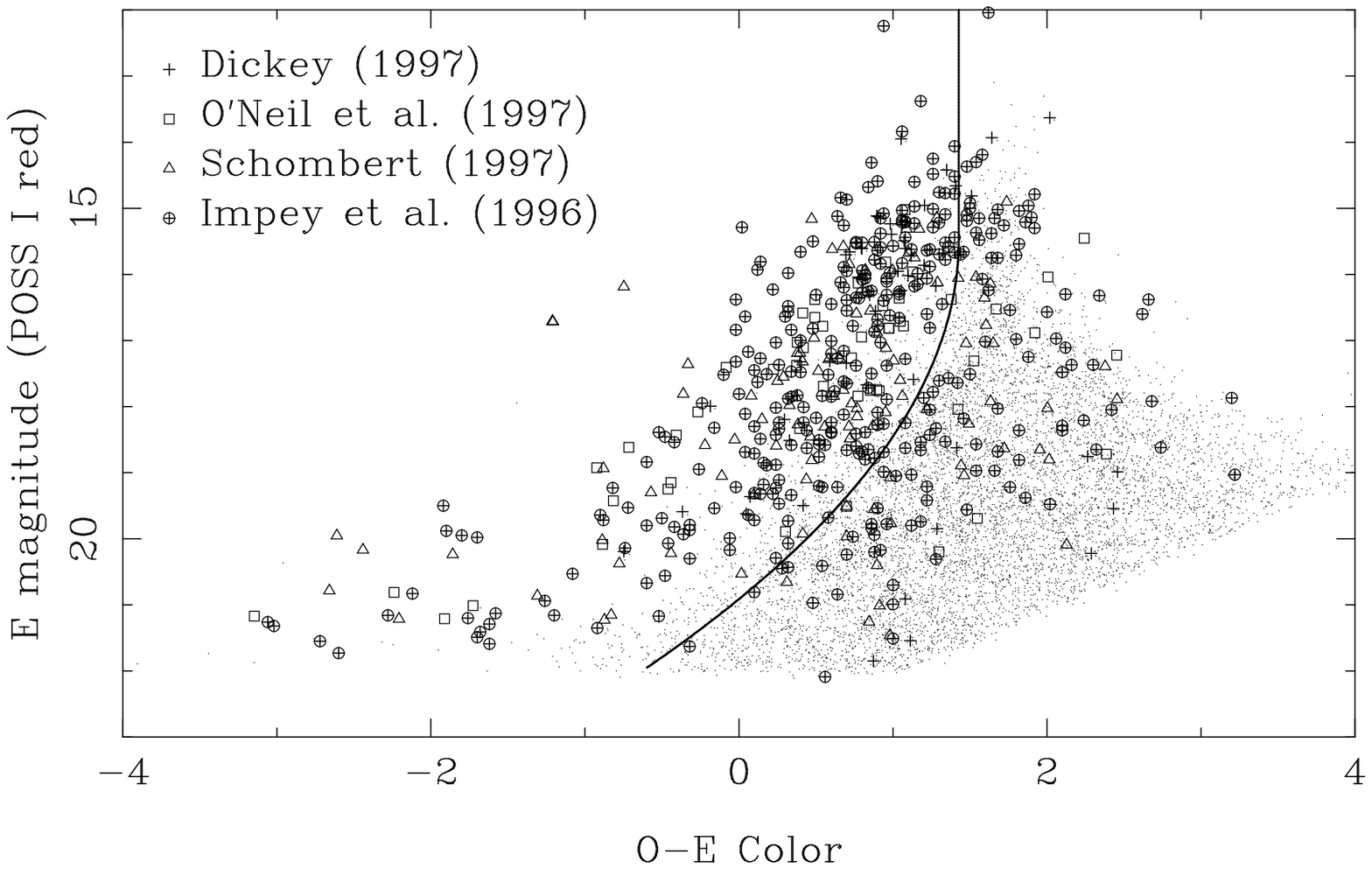]{The POSS I color-magnitude distribution of the APS
Catalog galaxies versus those of the {\HI}-rich and LSB galaxies in
\citet{dic97}, \citet{one97a}, \citet{sch97}, and \citet{imp96}.
Most of the {\HI}-rich and LSB galaxies lie blueward the blue edge boundary
defined by the bluest 10\% of APS Catalog galaxies.\label{ALLcolormag}}

\figcaption[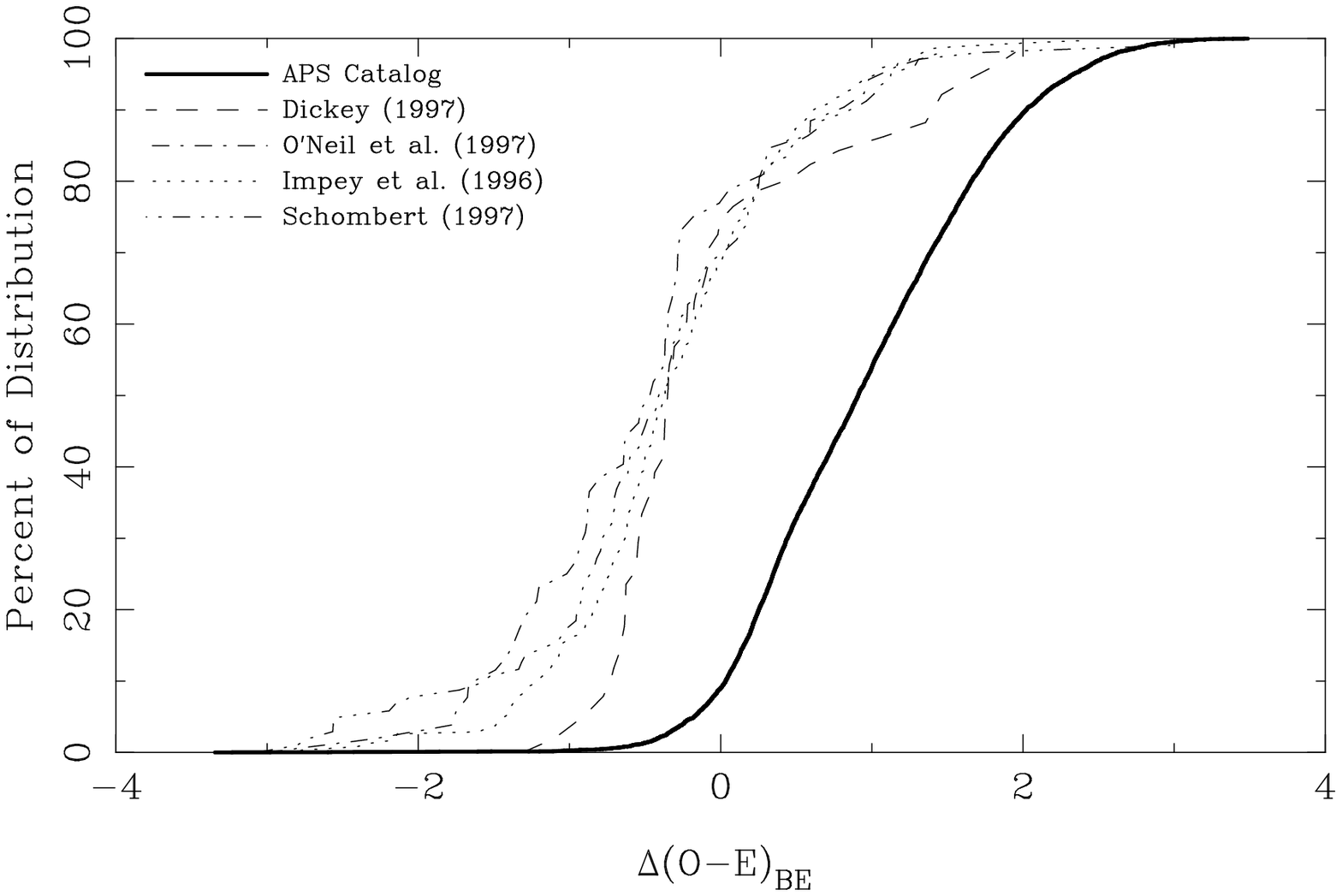]{The cumulative $\Delta(O-E)_{BE}$ distributions
for APS galaxies and various {\HI}-rich and LSB galaxy catalogs.  While (by
definition) only 10\% of APS galaxies lie blueward of the ``blue edge,'' 
well over 60\% of the {\HI}-rich and LSB galaxies lie blueward of
the ``blue edge.'' \label{BEdistribution}}

\figcaption[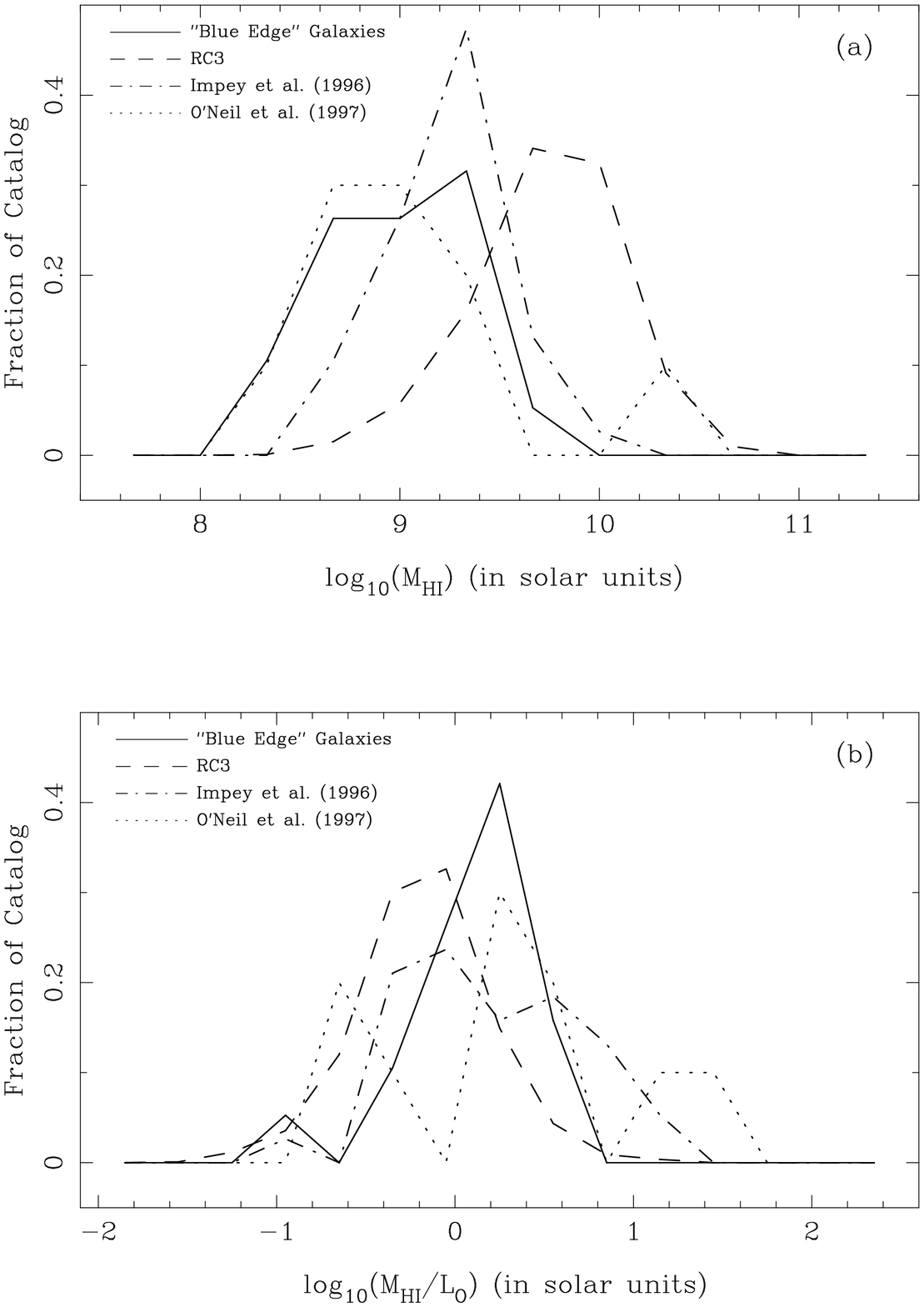]{(a) A plot of distribution of $M_{\HI}$ values in 4
galaxy catalogs with an imposed flux limit of $m_E<20$ and redshift range of
$4000$ {\kms}$<cz<9000$ {\kms}. The distributions are all scaled to the total
number of galaxies in the sample, with fractional counts computed in bins of
0.33 dex in $M_{\HI}$.  (b) A plot of distribution of $M_{\HI}/L_{O}$ values for
the same galaxy catalogs as in (a) computed for bin widths of 0.30 dex in
$M_{\HI}/L_{O}$. \label{HIdistros}}

\figcaption[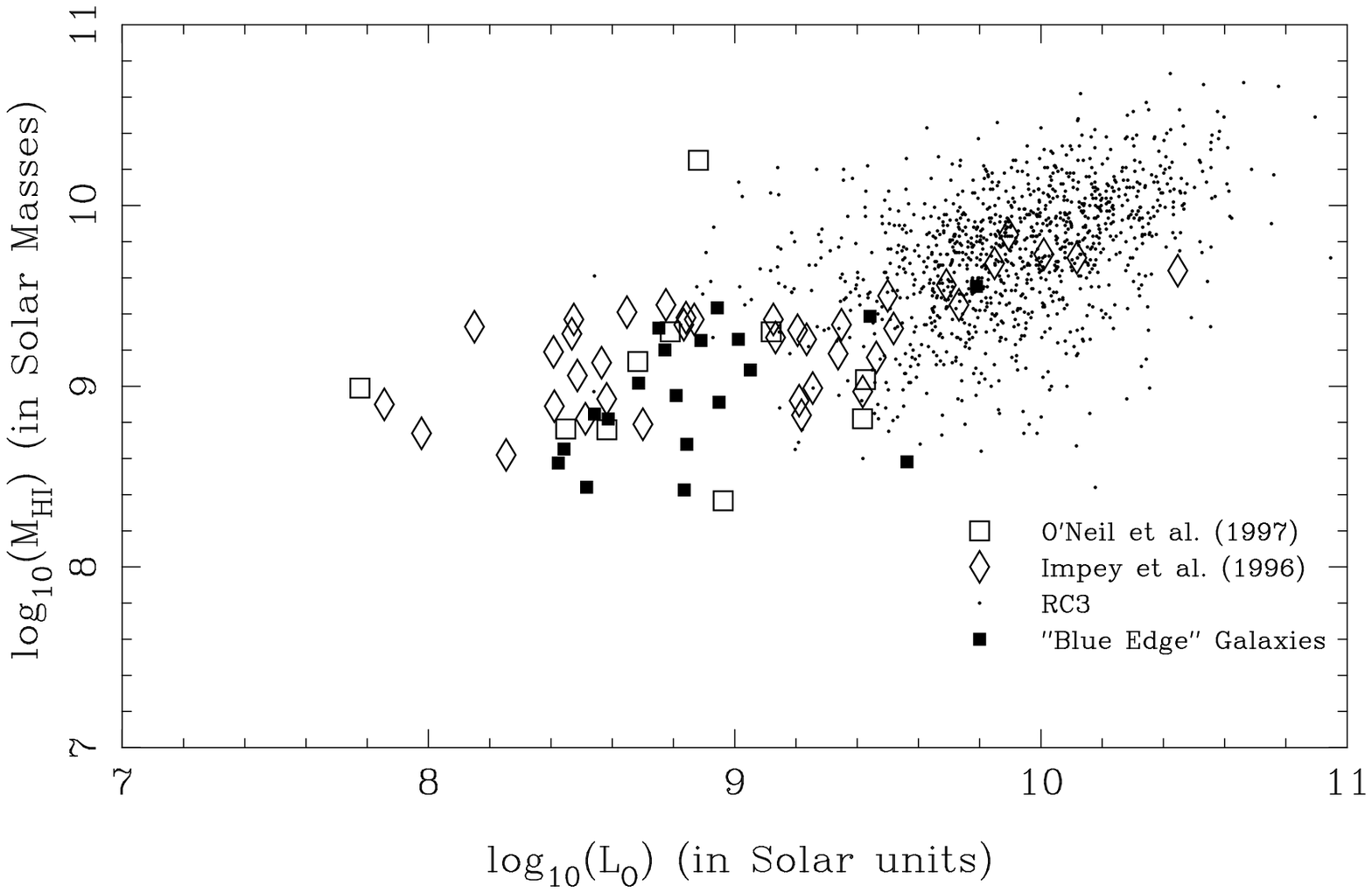]{A plot of $M_{\HI}$ versus $L_{O}$ (in solar
units) for the blue edge galaxies and other galaxy catalogs mentioned in
text, where $4000$ {\kms}$<cz<9000$ {\kms} and $E<20$. The imposed flux limit of
$m_E<20$ corresponds to a limiting luminosity of $\log(L_O) \gtrsim 7.7$ in
solar units, roughly the luminosity of the dimmest LSBs on this plot but
considerably dimmer than the least luminous blue edge galaxy. \label{lumHI}}

\figcaption[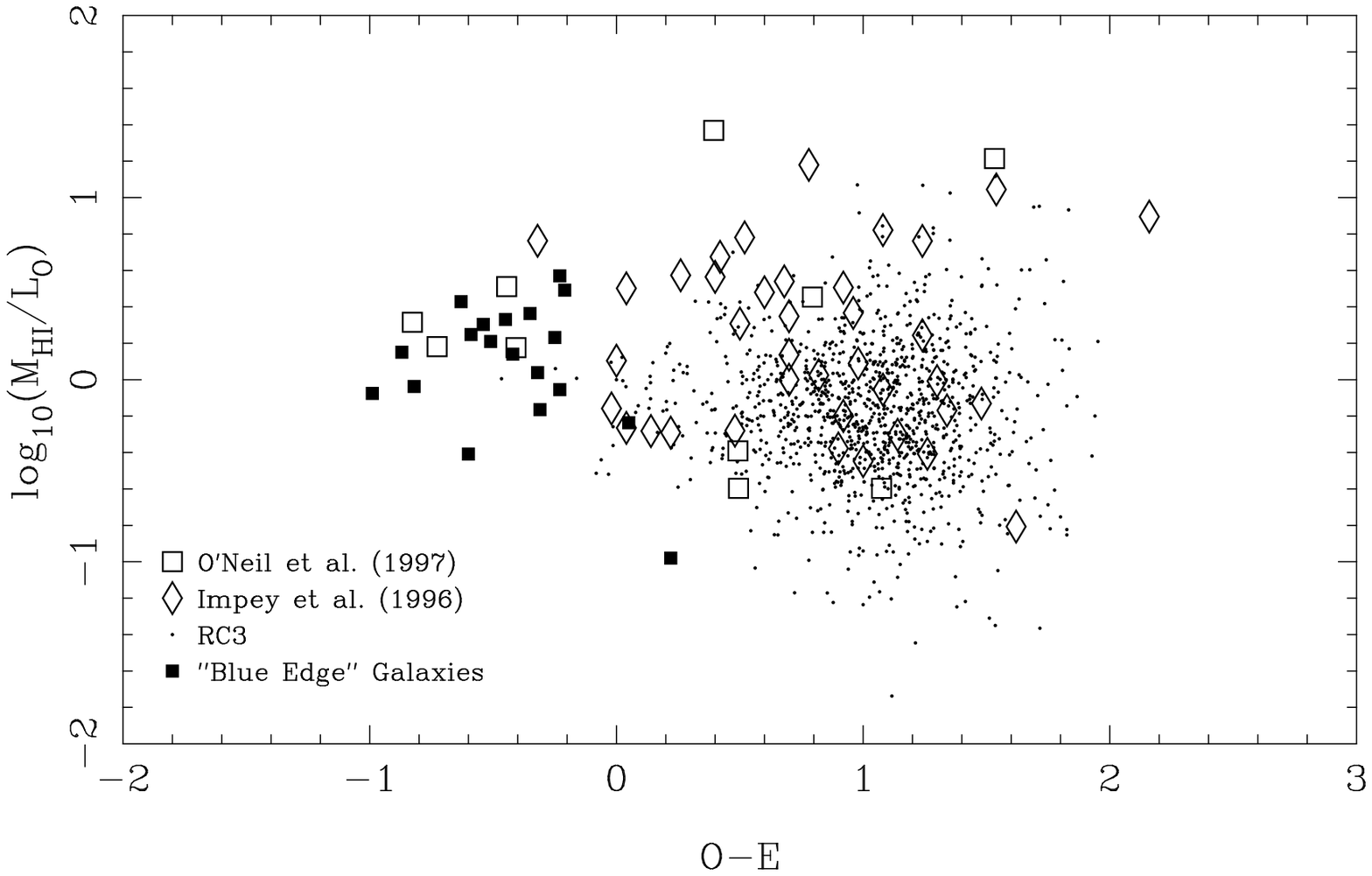]{A plot of $M_{\HI}/L_{O}$ (in solar units) versus
O-E color for the blue edge galaxies, the HSB galaxies from the RC3, and LSB
galaxies from \citet{imp96} and \citet{one97a}.  Whereas the bluest galaxies
only range over $\sim0.6$ dex in $M_{\HI}/L_{O}$, the reddest galaxies range
over 3 orders of magnitude in $M_{\HI}/L_{O}$, reaching gas mass-to-light ratios
an order of magnitude higher than those seen for the bluest galaxies.
\label{colorM2Lall}}

\figcaption[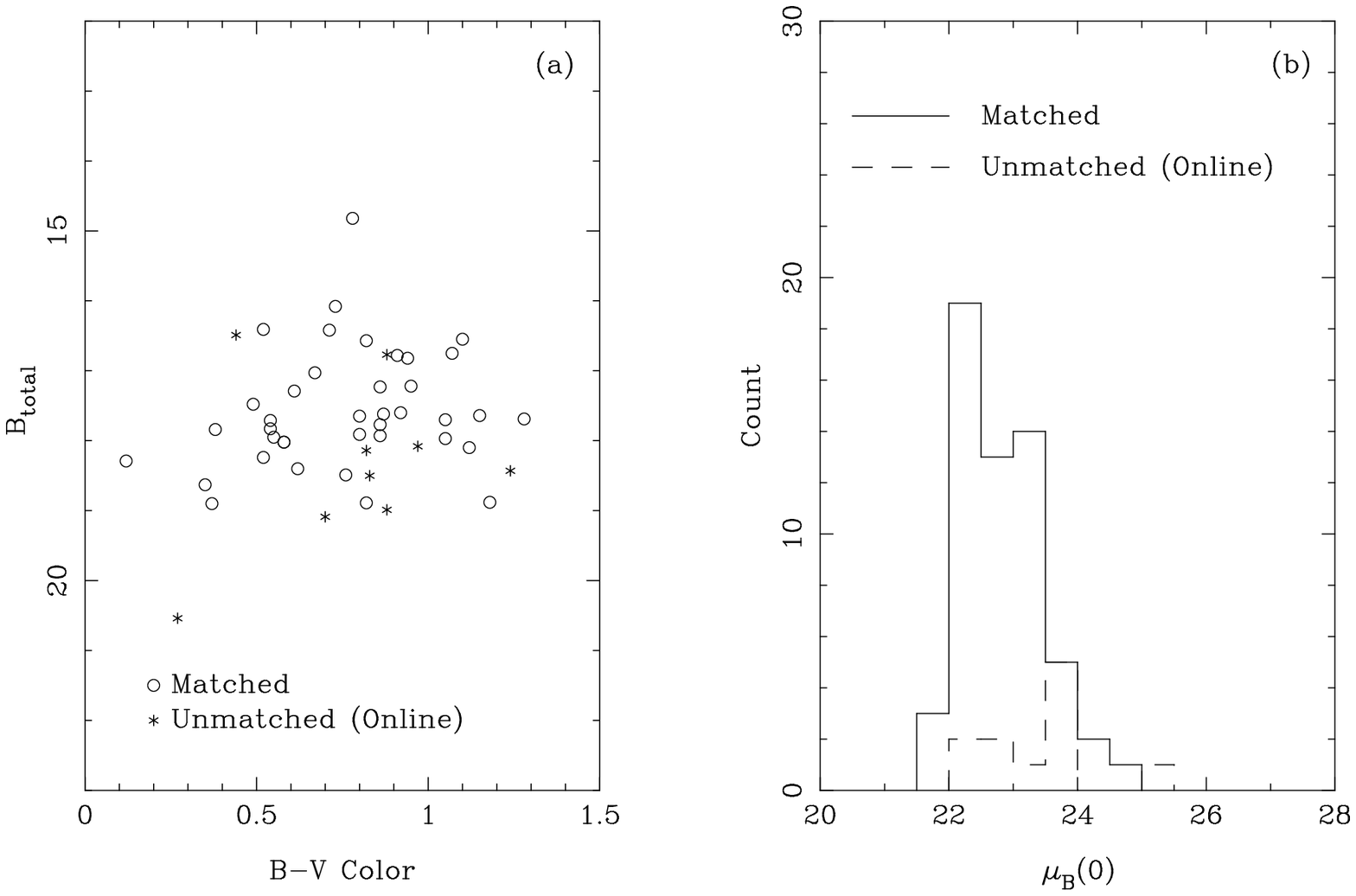]{(a) $B_{tot}$ vs. $B-V$ color-magnitude
diagrams for the OBC for galaxies matched or unmatched to the APS Catalog. 
(b) Histograms of the $B$ central surface brightnesses of OBC galaxies
(uncorrected for galaxy inclination) separated by whether the data is matched or
unmatched to the APS Catalog. There is evidence of the expected bias against
selecting the extremely low surface brightness objects from the OBC.
\label{distrocomp}}

\figcaption[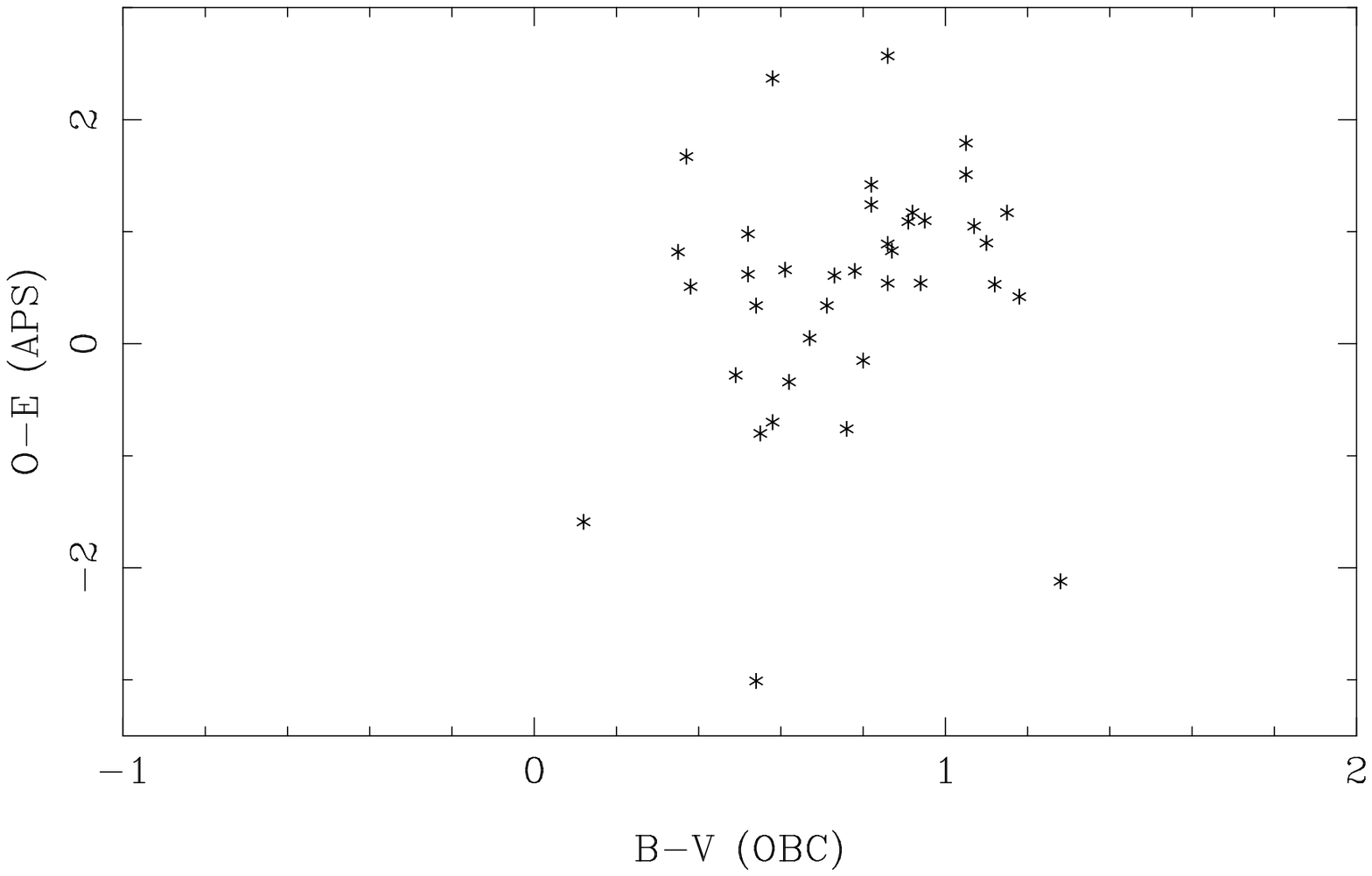]{A plot of $B-V$ color (from OBC) versus $O-E$
color (from the APS Catalog of the POSS I) for cross-identified OBC galaxies.
The two colors appear to be very weakly correlated.\label{colorcomp}}

\figcaption[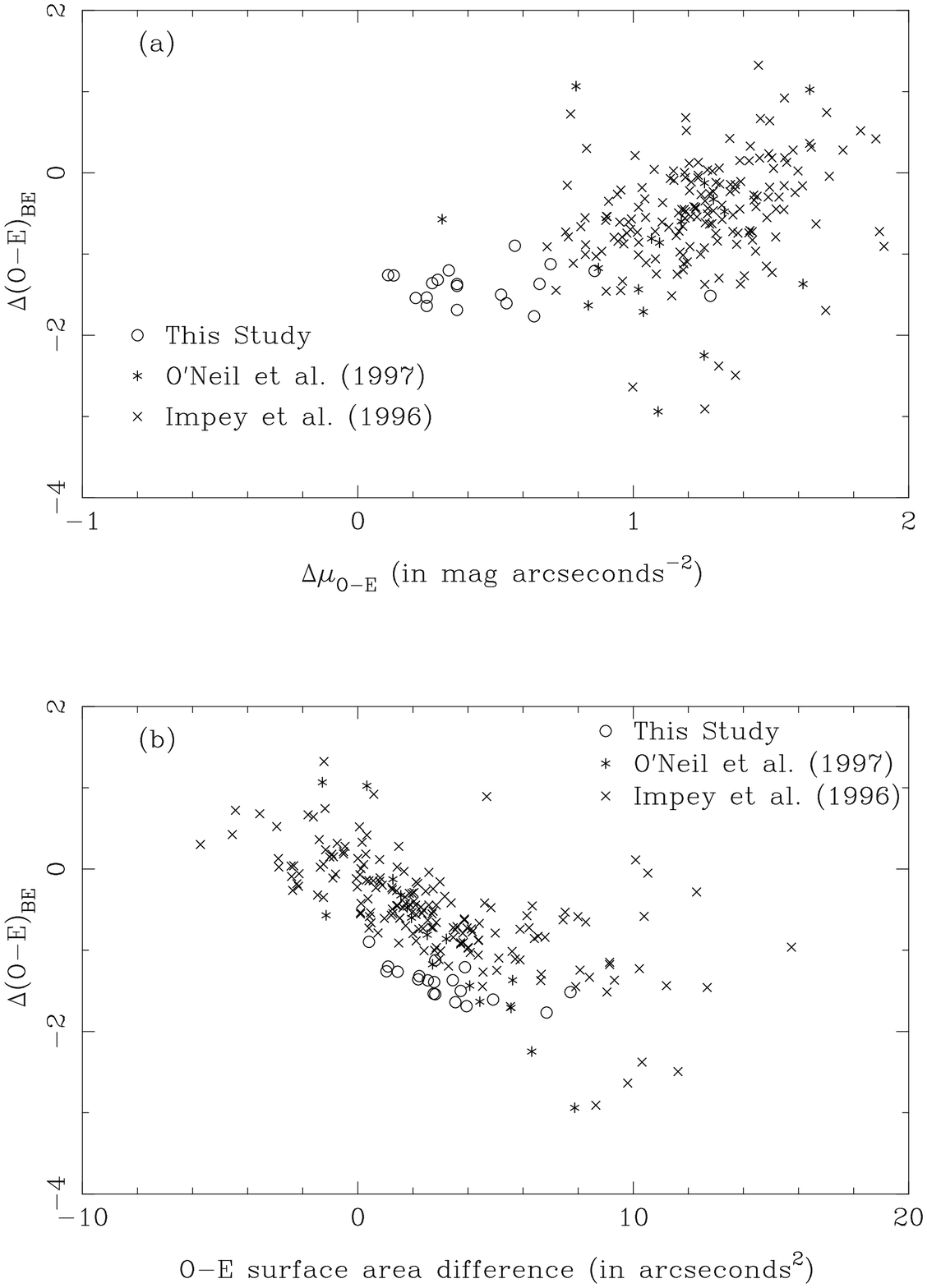]{Plots of the $O-E$ colors for galaxies in
\citet{imp96}, \citet{one97a}, and this study versus (a) the difference in $O$
and $E$ surface brightness and (b) the difference in $O$ and $E$ image surface
area.  \label{colorVparam}}

\figcaption[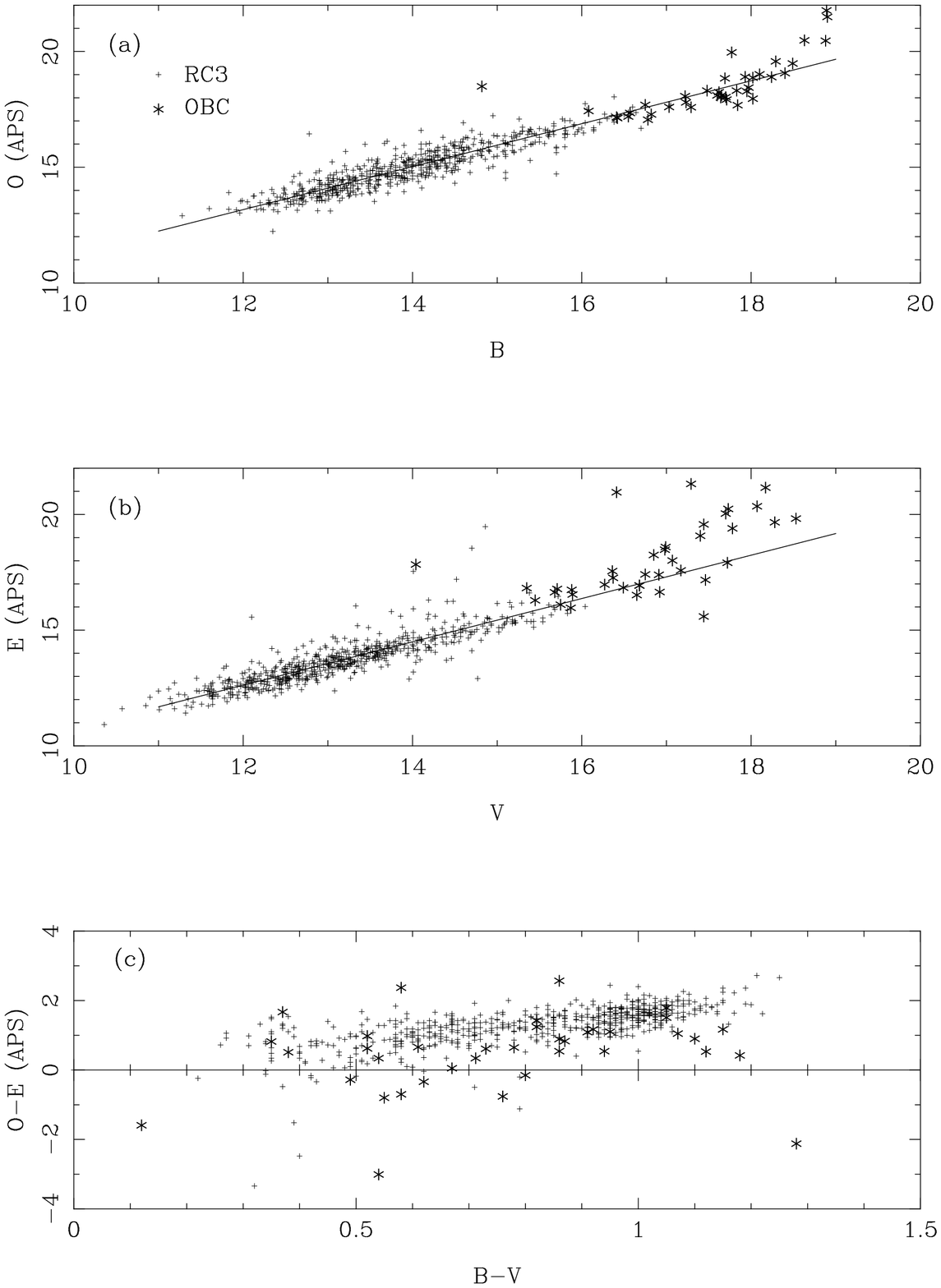]{Plots comparing the measured fluxes of RC3 versus
OBC galaxies. (a) A plot of POSS I $O$ magnitude versus $B$ magnitudes for both
the RC3 and OBC.  The best fit linear relationship for $O$ vs $B$, derived for
the RC3 galaxies alone, is show  The faintest OBC galaxies all lie above the
best fit line, suggesting their POSS I $O$ magnitudes are all fainter than the
RC3-based relationship between $O$ and $B$ would suggest.  (b) A similar plot to
(a) except for POSS I $E$ versus $V$ magnitudes.  In this case, the vast
majority of the OBC galaxies lie at fainter POSS I $E$ magnitudes than the
RC3-based relationship would suggest. (c) The combination of the two effects
illustrated in (a) and (b) is that the OBC galaxies have systematically bluer
POSS I $O-E$ colors than their RC3 counterparts for comparable $B-V$ colors.
\label{magcomparison}}

\figcaption[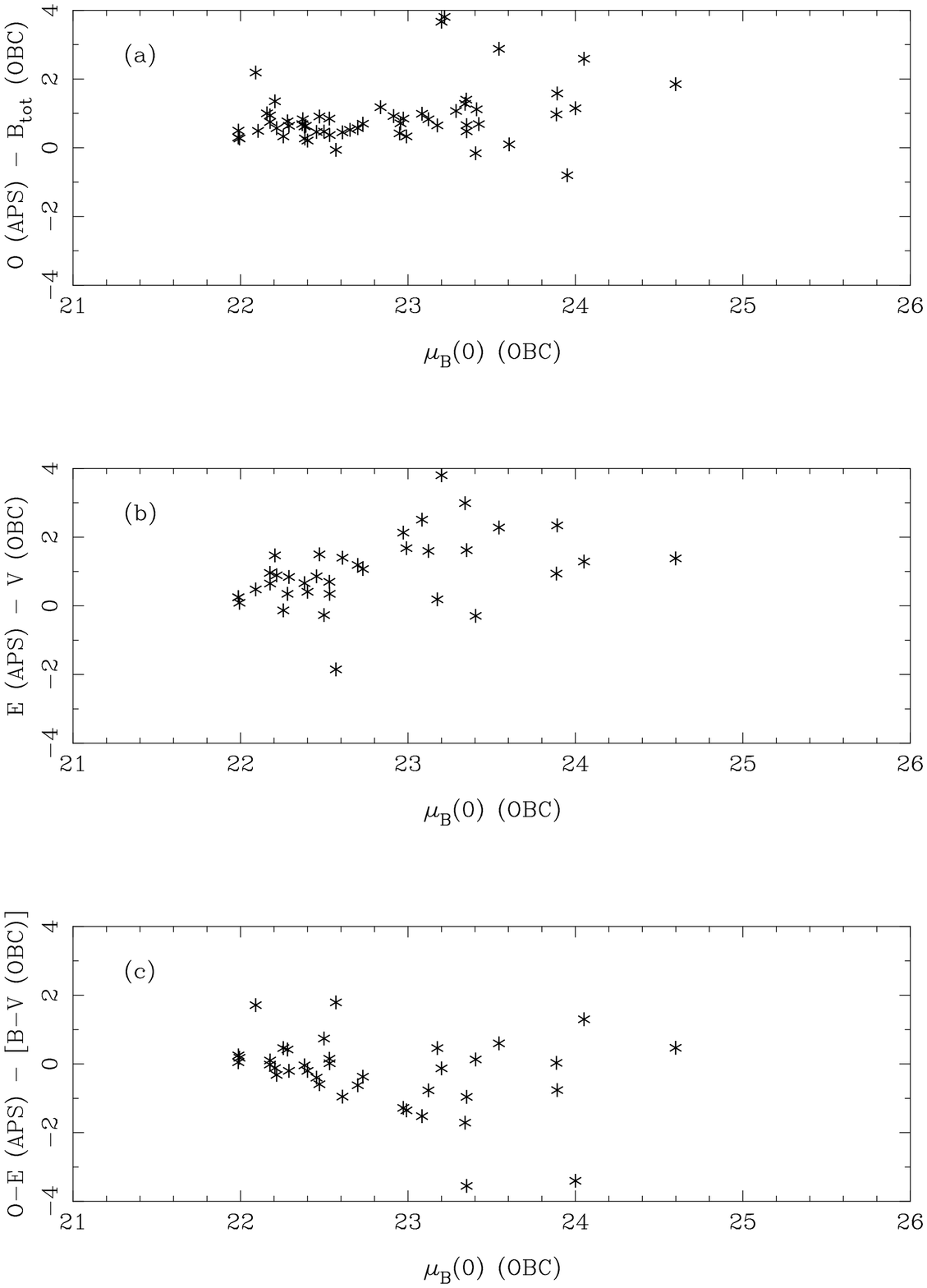]{(a) The difference of POSS I $O$ magnitude and $B$
magnitudes for OBC galaxies versus their central surface brightness
($\mu_{B}(0)$ from OBC). The best fit line to this data has a slope of
$0.383\pm0.171$, indicating a weak trend in $O-B$ versus $\mu_{B}(0)$.  (b) A
similar plot to (a) except for the difference in POSS I $E$ and $V$ magnitudes
versus $\mu_{B}(0)$. The effect is more dramatic in this case, with the best-fit
line on this plot having a slope of $0.984\pm0.266$. (c) The two effects
illustrated in (a) and (b) combine such that the difference between $O-E$ and
$B-V$ colors for OBC will become larger (with bluer $O-E$ than $B-V$ values) as
we move toward more diffuse $\mu_{B}(0)$ values. \label{magVmu}}

\clearpage
\begin{deluxetable}{lllll}
\tabletypesize{\scriptsize}
\tablecaption{POSS I Plate Parameters from the APS Catalog for POSS I Fields in this Study \label{tblflds}}
\tablewidth{0pt}
\tablehead{
\colhead{Field Number} & \multicolumn{2}{c}{Field Center} & \multicolumn{2}{c}{$\mu_{threshold}$} \\ 
 \cline{2-3} \cline{4-5} \\
 & \colhead{$\alpha$ (B1950)} & \colhead{$\delta$ (B1950)} & \colhead{O} & \colhead{E}
}
\startdata
P293 & 00:04:53.00 & +30:31:45.00 & 23.8 & 23.6 \\
P294 & 00:31:01.00 & +30:31:31.00 & 24.6 & 23.1 \\
P295 & 00:57:11.00 & +30:30:46.00 & 23.7 & 23.0 \\
P296 & 01:23:30.00 & +30:28:12.00 & 24.6 & 22.6 \\
P297 & 01:49:25.00 & +30:28:25.00 & 23.9 & 22.9 \\
P445 & 16:04:17.00 & +17:44:28.00 & 23.7 & 23.2 \\
\enddata
\end{deluxetable}

\clearpage

\begin{deluxetable}{crrrrrrl}
\tabletypesize{\scriptsize}
\tablecaption{{\HI}-rich Candidates List \label{tbl1}}
\tablewidth{470pt}
\tablehead{
\colhead{Object} & \colhead{$\alpha$ (B1950)}   & \colhead{$\delta$ (B1950)} &
\colhead{E\tablenotemark{a}} & \colhead{O-E\tablenotemark{a}}  & \colhead{E(B-V)\tablenotemark{b}} 
& \colhead{Date\tablenotemark{c}} & \colhead{Comments\tablenotemark{d}}
}
\startdata
MAPS-P293-137994 & 23:51:14.72 & +30:55:51.42 & 18.02 & -0.69 & 0.079 & 11 & Non-Detection \\
MAPS-P293-268610 & 00:02:39.36 & +28:36:23.43 & 18.74 & -0.21 & 0.052 & 08 \& 09 & Off Beam on\\
                &             &              &              &              &       &        & UGC 105 (8${th}$)\\
MAPS-P293-100987 & 00:04:52.45 & +31:47:35.03 & 17.76 & -0.59 & 0.040 & 16 & EOSD F409-05 (On Beam)\\*
                &             &              &              &              &       &    & Off Beam Detection\\
MAPS-P293-211328 & 00:09:59.12 & +29:40:06.63 & 19.19 & -0.84 & 0.054 & \nodata  & \\
MAPS-P293-249588 & 00:11:59.58 & +29:07:58.17 & 18.45 & -0.31 & 0.041 & \nodata  & \nodata  \\
MAPS-P293-249758 & 00:12:36.01 & +28:58:22.68 & 19.20 & -0.46 & 0.035 & 10 \& 11 & Strong RFI(10$^{th}$) \\*
                &             &              &              &              &       &          & Narrow RFI(11$^{th}$)\\
MAPS-P293-179271 & 00:13:36.89 & +30:13:54.80 & 18.28 & -0.33 & 0.064 & 20 & \nodata  \\
MAPS-P293-180077 & 00:16:18.49 & +30:14:11.05 & 16.79 & -0.01 & 0.075 & \nodata  & SRGb062.66 \\
MAPS-P293-239452 & 00:18:11.89 & +29:08:00.91 & 17.95 & -0.18 & 0.035 & \nodata  & \nodata  \\
MAPS-P294-766044 & 00:30:34.46 & +28:07:38.66 & 18.61 & -0.29 & 0.045 & 08 & Non-Detection \\
MAPS-P294-241809 & 00:31:15.32 & +31:10:32.02 & 15.89 & 0.29 & 0.056 & \nodata  & UGC 334 \\
MAPS-P294-727319 & 00:41:23.79 & +28:36:05.96 & 17.81 & -0.35 & 0.046 & 16 & UGC 467 \\
MAPS-P295-1977899 & 00:43:43.40 & +29:21:04.69 & 15.90 & 0.06 & 0.053 & \nodata  & CGCG 500-083 \\
MAPS-P295-2259922 & 00:43:51.97 & +28:44:55.64 & 19.75 & -0.99 & 0.061 & 10 \& 11 & Non-Detection \\*
                &             &              &              &              &       &          & Strong RFI(10$^{th}$) \\
MAPS-P295-922471 & 00:44:49.60 & +31:23:00.05 & 16.06 & 0.14 & 0.067 & \nodata  & CGCG 500-095 \\
MAPS-P295-2526723 & 00:45:28.88 & +28:13:59.64 & 18.16 & -0.32 & 0.069 & \nodata  & \nodata  \\
MAPS-P295-1880836 & 00:45:30.18 & +29:40:43.94 & 19.45 & -0.75 & 0.063 & 15 & Non-Detection \\
MAPS-P295-2261674 & 00:46:15.89 & +28:47:25.55 & 17.02 & -0.14 & 0.056 & \nodata  & \nodata  \\
MAPS-P295-1031187 & 00:47:26.93 & +31:13:33.74 & 17.82 & -0.22 & 0.063 & \nodata  & \nodata  \\
MAPS-P295-2078102 & 00:47:28.05 & +29:11:06.88 & 17.83 & -0.20 & 0.054 & \nodata  & \nodata  \\
MAPS-P295-2262929 & 00:47:52.74 & +28:50:56.86 & 16.63 & -0.14 & 0.058 & \nodata  & \nodata  \\
MAPS-P295-1577104 & 00:48:04.78 & +30:15:29.53 & 18.01 & -0.34 & 0.072 & 12 & FGC012A (On Beam)\\*
                &             &              &              &              &       &    & Non-Detection\\
                &             &              &              &              &       &    & Off Beam Detection\\
MAPS-P295-1474797 & 00:49:15.50 & +30:26:37.99 & 18.98 & -0.68 & 0.072 & 14 & Non-Detection  \\
MAPS-P295-585461 & 00:49:18.24 & +32:09:13.14 & 18.40 & -0.22 & 0.064 & 08 & Non-Detection  \\
MAPS-P295-699736 & 00:49:56.50 & +31:56:30.93 & 18.42 & -0.63 & 0.068 & 10 \& 11 &  \nodata  \\
MAPS-P295-1369071 & 00:50:18.47 & +30:44:33.45 & 18.51 & -0.23 & 0.055 & 12, 13, & NPM1G +30.0027 in \\*
                &             &              &              &              &       & 14, \& 15 & field (see text) \\
MAPS-P295-487139 & 00:50:38.45 & +32:21:58.70 & 17.36 & -0.23 & 0.078 & \nodata  & \nodata  \\
MAPS-P295-1790367 & 00:50:39.52 & +29:48:48.71 & 18.93 & -0.54 & 0.066 & 08 \& 09 & Non-Detection  \\
MAPS-P295-2270262 & 00:51:13.28 & +28:50:59.67 & 16.24 & 0.12 & 0.055 & \nodata  & \nodata  \\
MAPS-P295-823940 & 00:51:56.55 & +31:43:18.81 & 18.43 & -0.60 & 0.061 & 08 \& 09 & Non-Detection \\*
                &             &              &              &              &       &    &  Strong RFI(8$^{th}$)  \\
MAPS-P295-1894424 & 00:52:47.08 & +29:37:56.13 & 19.20 & -0.49 & 0.065 & \nodata  & \nodata  \\
MAPS-P295-1693362 & 00:52:49.86 & +30:07:37.67 & 15.87 & 0.05 & 0.065 & 08 & CGCG 501-047 \\
MAPS-P295-829092 & 00:53:34.38 & +31:41:46.88 & 19.03 & -0.58 & 0.063 & \nodata  & \nodata  \\
MAPS-P295-391670 & 00:54:34.29 & +32:35:19.01 & 17.68 & -0.14 & 0.058 & \nodata  & \nodata  \\
MAPS-P295-392980 & 00:54:54.66 & +32:29:53.83 & 16.47 & -0.37 & 0.056 & \nodata  & In MAPS-PP \\
MAPS-P295-500031 & 00:55:00.42 & +32:16:35.34 & 15.98 & -0.05 & 0.057 & \nodata  & CGCG 501-080 \\
MAPS-P295-296064 & 00:55:38.66 & +32:38:37.09 & 19.35 & -0.65 & 0.057 & 10 & Non-Detection  \\
MAPS-P295-951462 & 00:56:09.78 & +31:28:52.05 & 18.09 & -0.36 & 0.062 & \nodata  & \nodata  \\
MAPS-P295-1275729 & 00:56:19.32 & +30:53:48.70 & 17.09 & -0.04 & 0.067 & \nodata  & \nodata  \\
MAPS-P295-1386371 & 00:57:06.42 & +30:42:58.33 & 18.70 & -0.25 & 0.065 & \nodata  & \nodata  \\
MAPS-P295-83632 & 00:57:21.92 & +33:18:18.40 & 19.06 & -0.51 & 0.058 & 13 & Some RFI \\
MAPS-P295-1387790 & 00:57:32.61 & +30:43:53.15 & 18.41 & -0.25 & 0.062 & \nodata  & \nodata  \\
MAPS-P295-2202485 & 00:57:44.62 & +29:03:54.80 & 19.06 & -0.60 & 0.073 & \nodata  & \nodata  \\
MAPS-P295-1915216 & 00:58:22.06 & +29:40:19.07 & 18.88 & -0.82 & 0.059 & 10 & UGC 630 in field \\*
                &             &              &              &              &       &    & (see text) \\
MAPS-P295-1915437 & 00:58:24.48 & +29:39:55.20 & 19.21 & -0.71 & 0.059 & \nodata  & \nodata \\
MAPS-P295-1284440 & 00:58:45.37 & +30:55:24.13 & 16.86 & 0.03 & 0.058 & \nodata  & \nodata  \\
MAPS-P295-1067425 & 00:59:25.98 & +31:15:16.14 & 18.60 & -0.20 & 0.054 & 10 & Non-Detection  \\
MAPS-P295-1069708 & 00:59:52.52 & +31:16:36.26 & 17.66 & -0.33 & 0.053 & \nodata  & \nodata  \\
MAPS-P295-230677 & 01:00:24.99 & +32:51:30.43 & 18.49 & -0.30 & 0.059 & 08 \& 17 & Non-Detection  \\
MAPS-P295-156408 & 01:00:38.44 & +33:11:53.30 & 19.79 & -0.91 & 0.066 & \nodata  & \nodata  \\
MAPS-P295-2124445 & 01:00:59.18 & +29:19:21.73 & 16.81 & -0.14 & 0.063 & \nodata  & \nodata  \\
MAPS-P295-532118 & 01:01:00.18 & +32:16:23.27 & 18.50 & -0.45 & 0.067 & 10 & Non-Detection  \\
MAPS-P295-1615585 & 01:01:01.86 & +30:18:09.08 & 18.71 & -0.50 & 0.056 & 11 & Non-Detection  \\
MAPS-P295-325936 & 01:01:13.58 & +32:41:43.41 & 16.35 & -0.38 & 0.058 & \nodata  & UGC 657 \\
MAPS-P295-1827388 & 01:01:38.08 & +29:54:52.26 & 18.61 & -0.32 & 0.067 & 11 & Non-Detection \\
MAPS-P295-434368 & 01:01:50.90 & +32:30:24.65 & 19.48 & -0.84 & 0.059 & \nodata  & \nodata  \\
MAPS-P295-642385 & 01:01:55.27 & +32:03:44.94 & 18.42 & -0.25 & 0.065 & 08 \& 09 & \nodata  \\
MAPS-P295-100980 & 01:03:09.63 & +33:15:13.49 & 16.89 & -0.25 & 0.055 & \nodata  & In MAPS-PP \\
MAPS-P295-102863 & 01:03:38.14 & +33:20:16.32 & 16.01 & 0.20 & 0.054 & \nodata  & KUG 103+333 \\
MAPS-P295-657010 & 01:04:17.94 & +32:07:20.98 & 16.39 & -0.04 & 0.064 & \nodata  & UGC 679 \\
MAPS-P295-1202937 & 01:04:19.57 & +31:07:44.99 & 18.66 & -0.42 & 0.056 & 15 & \nodata  \\
MAPS-P295-345620 & 01:04:30.12 & +32:46:51.56 & 17.12 & -0.10 & 0.061 & \nodata  & \nodata\\
MAPS-P295-773634 & 01:05:02.09 & +31:57:13.89 & 19.54 & -0.87 & 0.057 & 10 & \nodata \\
MAPS-P295-1326244 & 01:06:28.46 & +30:49:28.58 & 15.84 & 0.21 & 0.064 & 13 & Non-Detection  \\
MAPS-P295-357095 & 01:07:14.10 & +32:38:17.48 & 16.47 & 0.07 & 0.064 & \nodata  & KUG0107+326A  \\
MAPS-P295-357226 & 01:07:16.59 & +32:39:56.58 & 17.95 & -0.34 & 0.064 & \nodata  & In MAPS-PP  \\
MAPS-P295-357773 & 01:07:25.88 & +32:38:05.88 & 19.19 & -0.70 & 0.061 & \nodata  & \nodata  \\
MAPS-P295-359427 & 01:07:57.89 & +32:45:57.00 & 16.44 & 0.13 & 0.052 & \nodata  & KUG0107+327 \\
MAPS-P295-362727 & 01:09:11.75 & +32:45:41.71 & 17.57 & -0.15 & 0.056 & \nodata  & KUG0107+327A \\
MAPS-P295-910484 & 01:09:14.09 & +31:41:16.90 & 17.03 & -0.23 & 0.059 & 14 & Strong RFI \\
MAPS-P295-576122 & 01:11:48.60 & +32:15:15.97 & 18.41 & -0.32 & 0.060 & 10, 11 & \nodata  \\*
               &             &              &              &              &       &  \& 17 & \\
\enddata
\tablenotetext{a}{These values are all corrected for Galactic extinction as described in the text.}
\tablenotetext{b}{E(B-V) from \citet{sch98}.}
\tablenotetext{c}{Observation dates (in August 1998) listed if observed.  All dates UTC.}
\tablenotetext{d}{All cross-identifications are within 0.5{{\arcmin}} of known position, and thus falling
well within the Arecibo beam.  ``MAPS-PP'' refers to the Minnesota Automated Plate Scanner Pisces-Perseus
Catalog of \citet{cab98}.}
\end{deluxetable}

\clearpage
\begin{deluxetable}{lrrrcccc}
\tabletypesize{\scriptsize}
\tablecaption{{\HI} Observations \label{tbl2}}
\tablewidth{0pt}
\tablehead{
\colhead{Object} & \colhead{$F_{obs}$} & \colhead{V\tablenotemark{a}} & 
\colhead{${\Delta}V$} & \colhead{Group} & \colhead{$\log M_{\HI}$} &  \colhead{$\log L$}  
& \colhead{$M_{\HI}/L_{O}$} \\
\colhead{}& \colhead{(mJy-{\kms})} & \colhead{({\kms})} 
& \colhead{({\kms})} & \colhead{} & \colhead{($M_{\sun}$)} & \colhead{($L_{\sun}$)} 
& \colhead{(${\sun}$)}
}
\startdata
MAPS-P293-268610 & 857.5 (22.3) & 8361 & 151 & 1 & 9.44 & 8.86 & 3.78\\
MAPS-P293-100987 & 1704.0 (23.4) & 4851 & 124 & 1 & 9.26 & 8.77 & 3.06\\
MAPS-P293-249758 & 483.7 (16.4) & 6890 & 116 & 1 & 9.02 & 8.50 & 3.27\\
MAPS-P293-179271 & 508.1 (25.1) & 7288 & 129 & 1 & 9.09 & 8.92 & 1.47\\
MAPS-P294-727319 & 1692.8 (19.2) & 4817 & 137 & 1 & 9.25 & 8.75 & 3.17\\
MAPS-P295-699736 & 1457.5 (14.1) & 4899 & 124 & 1 & 9.20 & 8.52 & 4.82\\
MAPS-P295-1790367 & 628.2 (11.8) & 4956 & 138 & 1 & 8.85 & 8.33 & 3.31\\
MAPS-P295-823940 & 204.2 (10.3) & 5362 & 61 & 1 & 8.43 & 8.59 & 0.68\\
MAPS-P295-1693362 & 1766.3 (21.5) & 6667 & 241 & 1 & 9.55 & 9.81 & 0.55\\
MAPS-P295-1915216 & 389.5 (19.0) & 6788 & 121 & 1 & 8.91 & 8.62 & 1.95\\
MAPS-P295-1915216 & 763.7 (16.7) & 4878 & 156 & 1 & 8.92 & 9.71 & 0.16\\
(UGC 630) & & & & & & & \\
MAPS-P295-642385 & 651.9 (14.6) & 4716 & 94 & 1 & 8.82 & 8.49 & 2.15\\
MAPS-P295-1202937 & 496.1 (10.7) & 6280 & 168 & 1 & 8.95 & 8.64 & 2.04\\
MAPS-P295-1326244 & 279.9 (13.7) & 5483 & 44 & 1 & 8.58 & 9.65 & 0.09\\
MAPS-P295-910484 & 1194.3 (43.9) & 6701 & 170 & 1 & 9.39 & 9.35 & 1.09\\
OPSPNT & 920.6 (21.5) & 5998 & 170 & 2 & \nodata & \nodata & \nodata\\
MAPS-P295-1369071 & 1286.7 (18.3) & 5995 & 175 & 2 & 9.32 & 8.66 & 4.58\\
MIDPNT & 794.3 (21.9) & 5987 & 162 & 2 & \nodata & \nodata & \nodata\\
NPM1G +30.0027 & 311.6 (28.6) & 5993 & 229 & 2 & 8.71 & 10.02 & 0.05\\
MAPS-P295-2259922 & 187.0 (12.9) & 5698 & 129 & 3 & 8.44 & 8.12 & 2.10\\
MAPS-P295-83632 & 436.7 (17.9) & 4757 & 74 & 3 & 8.65 & 8.24 & 2.59\\
MAPS-P295-773634 & 342.0 (28.7) & 4918 & 221 & 3 & 8.58 & 8.08 & 3.16\\
MAPS-P295-576122 & 232.9 (13.9) & 6714 & 135 & 3 & 8.68 & 8.72 & 0.91\\
UGC 105 & 793.6 (28.2) & 8065 & 233 & 4 & $>9.37$ & 10.47 & $>0.08$\\
MAPS-P293-102675 & 1859.8 (24.4) & 6570 & 240 & 4 & $>9.56$ & 9.65 & $>0.82$\\
MAPS-P294-444433 & 290.9 (23.7) & 6697 & 129 & 4 & $>8.77$ & 9.01 & $>0.58$\\
NGC 634 & 4771.5 (38.0) & 4928 & 512 & 5 & 9.72 & 10.49 & 0.17\\
\enddata
\tablenotetext{a}{All velocities are heliocentric. Comparison between velocities
determined using flux density thresholds of 50\% the mean flux density and 20\%
the maximum flux density showed the difference between these two methods to
always be less than the uncertainty due to the channel width, therefore we take
the uncertainty to be 2 channel widths or 10 {\kms} in all cases.}
\end{deluxetable}

\clearpage

\begin{deluxetable}{lrrrr}
\tabletypesize{\scriptsize}
\tablecaption{Comparison Galaxy Catalogs Used \label{CompCats}}
\tablewidth{0pt}
\tablehead{
\colhead{} & \multicolumn{4}{c}{Number of Catalogs Galaxies used for comparison of...} \\
 \cline{2-5} \\
Source & BBD & Color-magnitude & $M_{\HI}$ & Photometry }
\startdata
ZCAT (February 2000) & 1648 & 1648 &   0  & 0 \\
\citet{dic97}        & 46   & 46   &   0  & 0 \\
\citet{imp96}        & 343  & 343  &  38  & 0 \\
\citet{one97a}       & 53   & 53   &  10  & 38 \\
\citet{sch97}        & 0    & 104  &   0  & 0 \\
RC3                  & 0    & 0    & 1126 & 617 \\
\enddata
\end{deluxetable}

\begin{deluxetable}{lrrrrrrr}
\tabletypesize{\scriptsize}
\tablecaption{Estimated Probabilities of Identical Parent $\Delta(O-E)_{BE}$ Distributions \label{probBE}}
\tablewidth{0pt}
\tablehead{
\colhead{} & \colhead{Sample Size} & \multicolumn{6}{c}{$\log$(probability)} \\
 \cline{3-8} \\
Comparison Sample &  & \colhead{(1)} & \colhead{(2)} & \colhead{(3)} & \colhead{(4)} & \colhead{(5)} & \colhead{(6)} }
\startdata
(1) APS & 7943 & 0.00 & -17.22 & -20.56 & -111.13 & -23.88 & -34.93\\
(2) Blue edge galaxies & 19 & -17.22 & 0.00 & -12.20 & -12.63 & -7.05 & -10.38\\
(3) \citet{dic97} & 51 & -20.56 & -12.20 & 0.00 & -0.64 & -2.20 & -1.51\\
(4) \citet{imp96} & 347 & -111.13 & -12.63 & -0.64 & 0.00 & -1.33 & -0.53\\
(5) \citet{one97a} & 52 & -23.88 & -7.05 & -2.20 & -1.33 & 0.00 & -0.39\\
(6) \citet{sch97} & 103 & -34.93 & -10.38 & -1.51 & -0.53 & -0.39 & 0.00\\
\enddata
\end{deluxetable}

\clearpage

\begin{deluxetable}{lrrrrr}
\tabletypesize{\scriptsize}
\tablecaption{Estimated Probabilities of Identical Parent $M_{\HI}$ Distributions \label{probM}}
\tablewidth{0pt}
\tablehead{
\colhead{} & \colhead{Sample Size} & \multicolumn{4}{c}{$\log$(probability)} \\
 \cline{3-6} \\
Comparison Sample & & \colhead{(1)} & \colhead{(2)} & \colhead{(3)} & \colhead{(4)} }
\startdata
(1) Blue edge galaxies & 19 & 0.00 & -1.10 & -0.06 & -9.50\\
(2) \citet{imp96} & 38 & -1.10 & 0.00 & -0.93 & -12.23\\
(3) \citet{one97a} & 10 & -0.06 & -0.93 & 0.00 & -5.40\\
(4) RC3 & 1126 & -9.50 & -12.23 & -5.40 & 0.00\\
\enddata
\end{deluxetable}

\begin{deluxetable}{lrrrrr}
\tabletypesize{\scriptsize}
\tablecaption{Estimated Probabilities of Identical Parent $M_{\HI}/L_{O}$ Distributions \label{probM2L}}
\tablewidth{0pt}
\tablehead{
\colhead{} & \colhead{Sample Size} & \multicolumn{4}{c}{$\log$(probability)} \\
 \cline{3-6} \\
Comparison Sample & & \colhead{(1)} & \colhead{(2)} & \colhead{(3)} & \colhead{(4)} }
\startdata
(1) blue edge Galaxies & 19 & 0.00 & -0.70 & -0.27 & -2.53\\
(2) \citet{imp96} & 38 & -0.70 & 0.00 & -0.12 & -4.10\\
(3) \citet{one97a} & 10 & -0.27 & -0.12 & 0.00 & -2.55\\
(4) RC3 & 1126 & -2.53 & -4.10 & -2.55 & 0.00\\
\enddata
\end{deluxetable}

\clearpage
\begin{figure} 
\plotone{Cabanela.fig01.eps}\\
Figure \ref{colormag}
\end{figure}
\clearpage

\begin{figure} 
\plotone{Cabanela.fig02.eps}\\
Figure \ref{cumilrfi}
\end{figure}
\clearpage

\begin{figure} 
\plotone{Cabanela.fig03.eps}\\
Figure \ref{group1}
\end{figure}
\clearpage

\begin{figure} 
\plotone{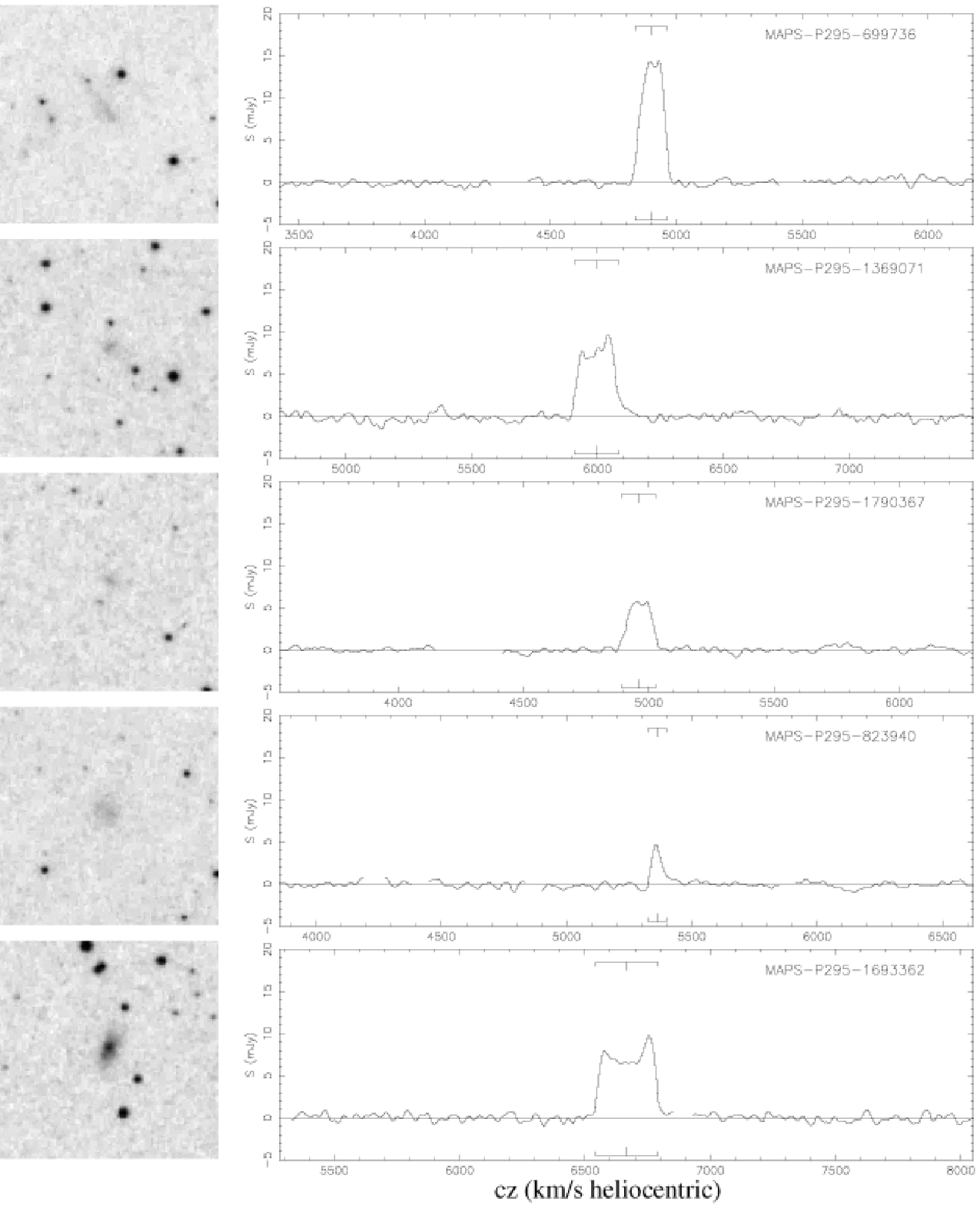}\\
Figure \ref{group1} (Continued)
\end{figure}
\clearpage

\begin{figure} 
\plotone{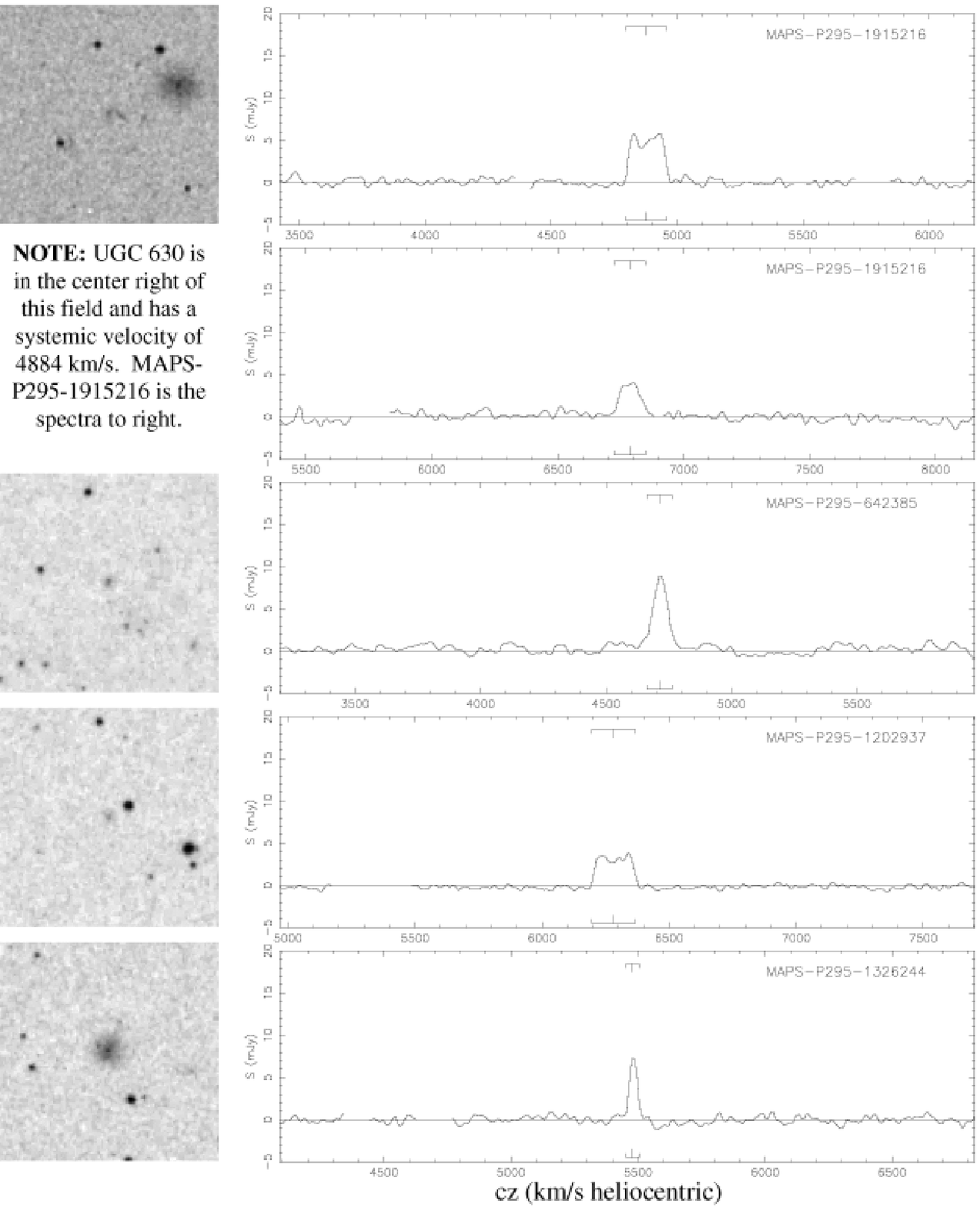}\\
Figure \ref{group1} (Continued)
\end{figure}
\clearpage

\begin{figure} 
\plotone{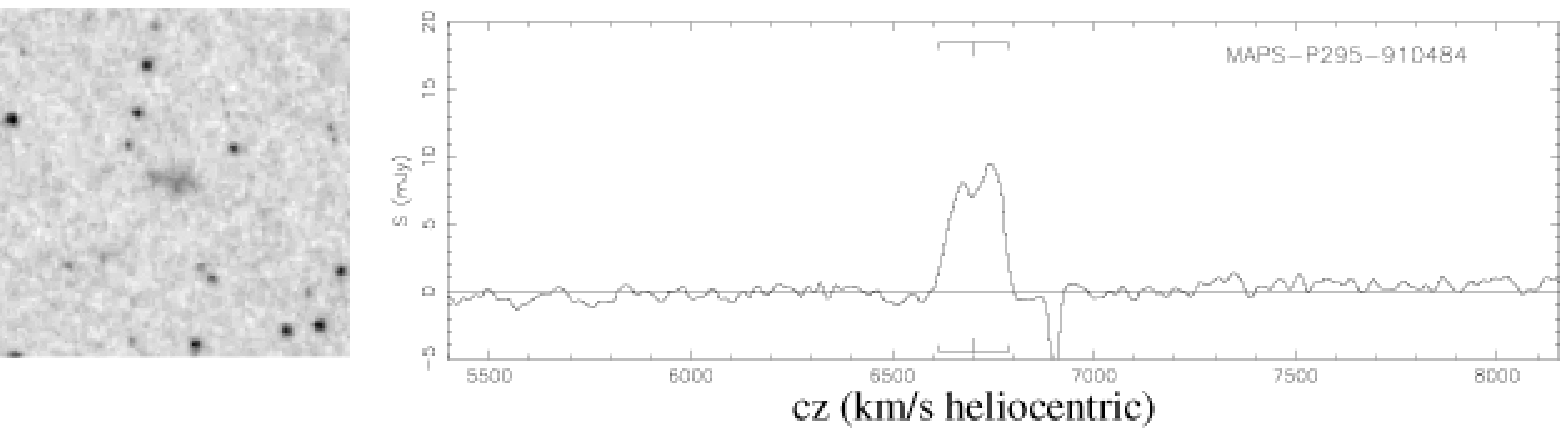}\\
Figure \ref{group1} (Continued)
\end{figure}
\clearpage

\begin{figure} 
\epsscale{0.6}
\plotone{Cabanela.fig04.eps}\\
Figure \ref{group2}
\end{figure}
\clearpage

\begin{figure} 
\epsscale{1}
\plotone{Cabanela.fig05.eps}\\
Figure \ref{group3}
\end{figure}
\clearpage

\begin{figure} 
\plotone{Cabanela.fig06.eps}\\
Figure \ref{group4}
\end{figure}
\clearpage

\begin{figure} 
\plotone{Cabanela.fig07.eps}\\
Figure \ref{ridgeHI}
\end{figure}
\clearpage

\begin{figure} 
\plotone{Cabanela.fig08.eps}\\
Figure \ref{ourbivar}
\end{figure}
\clearpage

\begin{figure} 
\plotone{Cabanela.fig09.eps}\\
Figure \ref{bivariate}
\end{figure}
\clearpage

\begin{figure} 
\plotone{Cabanela.fig10.eps}\\
Figure \ref{ALLcolormag}
\end{figure}
\clearpage

\begin{figure} 
\plotone{Cabanela.fig11.eps}\\
Figure \ref{BEdistribution}
\end{figure}
\clearpage

\begin{figure} 
\epsscale{0.8}
\plotone{Cabanela.fig12.eps}\\
Figure \ref{HIdistros}
\end{figure}
\clearpage

\begin{figure} 
\plotone{Cabanela.fig13.eps}\\
Figure \ref{lumHI}
\end{figure}
\clearpage

\begin{figure} 
\plotone{Cabanela.fig14.eps}\\
Figure \ref{colorM2Lall}
\end{figure}
\clearpage

\begin{figure} 
\plotone{Cabanela.fig15.eps}\\
Figure \ref{distrocomp}
\end{figure}
\clearpage

\begin{figure} 
\plotone{Cabanela.fig16.eps}\\
Figure \ref{colorcomp}
\end{figure}
\clearpage

\begin{figure} 
\epsscale{0.8}
\plotone{Cabanela.fig17.eps}\\
Figure \ref{colorVparam}
\end{figure}
\clearpage

\begin{figure}
\epsscale{0.8}
\plotone{Cabanela.fig18.eps}\\ 
Figure \ref{magcomparison}
\end{figure}
\clearpage

\begin{figure} 
\epsscale{0.8}
\plotone{Cabanela.fig19.eps}\\
Figure \ref{magVmu}
\end{figure}
\clearpage



\begin{thebibliography}{}
\bibitem[Briggs (1997)]{bri97} Briggs, F.H. 1997, \apj, 484, 618
\bibitem[Burstein and Heiles (1982)]{bur82} Burstein, D. and Heiles, C. 1982, \aj, 87, 1165
\bibitem[Bothun (1982)]{bot82} Bothun, G.D. 1982, \apjs, 50, 39
\bibitem[Bothun (1984)]{bot84} Bothun, G.D. 1984, \apj, 277, 532
\bibitem[Bothun, Impey, Malin, \& Mould(1987)]{bot87} Bothun, G.D., Impey, C.D., Malin, D.F., \& Mould, J.R.\ 1987, \aj, 94, 23
\bibitem[Bothun {\etal} (1997)]{bot97} Bothun, G.D., Impey, C., and McGaugh, S. 1997, \pasp, 109, 745
\bibitem[Cabanela and Aldering (1998)]{cab98} Cabanela, J.E., and Aldering, G. 1998, \aj, 116, 1094
\bibitem[Cabanela and Dickey (1999)]{cab99} Cabanela, J.E., and Dickey, J.M. 1999, \aj, 118, 46
\bibitem[Cardelli, Clayton, \& Mathis (1989)]{car89} Cardelli, J. A., Clayton, G. C, and Mathis, J. S. 1989, \apj, 345, 245
\bibitem[Cayatte, Kotanyi, Balkowski, \& van Gorkom(1994)]{cay94} Cayatte, V., Kotanyi, C., Balkowski, C., \& van Gorkom, J.H.\ 1994, \aj, 107, 1003
\bibitem[de Blok, McGaugh, and van der Hulst (1996)]{dBl96} de Blok, W.J.G., McGaugh, S.S., and van der Hulst, J.M. 1996, \mnras, 283, 18
\bibitem[de Vaucouleurs {\etal} (1991)]{deV91} de Vaucouleurs, G., de Vaucouleurs, A., Corwin, H.G. Jr., Buta, R.J., Paturel, G., and Fouqu{\'{e}}, P.  1991, {\it Third Reference Catalogue of Bright Galaxies} (Springer-Verlag, New York)
\bibitem[Dickey (1997)]{dic97} Dickey, J.M. 1997, \aj, 113, 1939
\bibitem[Disney (1976)]{dis76} Disney, M.J. 1976, Nature, 263, 573
\bibitem[Dressler (1980)]{dre80} Dressler, A. 1980, \apj, 236, 351
\bibitem[Driver and Cross (2000)]{dri00} Driver, S. and Cross, N. 2000, in {\it Mapping the Hidden Universe (ASP Conference Series Vol. 218)}, (Astronomical Society of the Pacific, San Francisco), 309.
\bibitem[Eder {\etal} (1989)]{ede89} Eder, J., Oemler, A.J., Schombert, J.M. and Dekel, A. 1989, \apj, 340, 29
\bibitem[Giovanelli and Haynes (1985)]{gio85} Giovanelli, R. and Haynes, M.P. 1985, \apj, 292, 404
\bibitem[Giovanelli, Haynes \& Chincarini (1986)]{gio86} Giovanelli, R., Haynes, M.P., and Chincarini, G.L. 1986, \apj, 300, 77
\bibitem[Giovanelli and Haynes (1988)]{gio88} Giovanelli, R. and Haynes, M.P. 1988, in {\it Large-Scale Structures of the Universe, IAU Symp. 130}, (Kluwer Academic Publishes, Dordrecht), 113
\bibitem[Giraud (1987)]{gir87} Giraud, E. 1987, \aap, 178, 310
\bibitem[Gerritsen and de Blok (1999)]{ger99} Gerritsen, J.P.E. and de Blok, W.J.G. 1999, \aap, 342, 655
\bibitem[Huchra {\etal} (1992)]{huc92} Huchra, J.,  Geller, M., Clemens, C., Tokarz, S and Michel, A. 1992, Bull. C.D.S. 41, 31. (ZCAT, February 2000 version)
\bibitem[Impey {\etal} (1996)]{imp96} Impey, C.D., Sprayberry, D., Irwin, M.J., and Bothun, G.D. 1996, \apjs, 105, 209
\bibitem[Impey and Bothun (1997)]{imp97} Impey, C. and Bothun, G.D. 1997, \araa, 35, 267
\bibitem[Lasker et al.(1990)]{las90} Lasker, B.M., Sturch, C.R., McLean, B.J., Russell, J.L., Jenkner, H., \& Shara, M.M.\ 1990, \aj, 99, 2019
\bibitem[Lavezzi and Dickey (1997)]{lav97} Lavezzi, T.E. and Dickey, J.M. 1997, \aj, 114, 2437
\bibitem[Maia, Willmer, \& Da Costa (1998)]{mai98} Maia, M.A.G., Willmer, C.N.A., and Da Costa, L.N. 1998, \aj, 115, 49
\bibitem[Martin and Kennicutt (2001)]{mar01} Martin, C.L. and Kennicutt, R.C. 2001, \apj, 555, 301
\bibitem[Matthews and Gallagher (1997)]{mat97} Matthews, L.D. and Gallagher, J.S. 1997, \aj, 114, 1899
\bibitem[Meurer {\etal} (1996)]{meu96} Meurer, G.R., Carignan, C., Beaulieu, S.F., and Freeman, K.C. 1996, \aj, 111, 1551
\bibitem[Odewahn {\etal} (1992)]{ode92} Odewahn, S.C., Stockwell, E.B., Pennington, R.L., Humphreys, R.M., and Zumach, W.A. 1992, \aj, 103, 318
\bibitem[O'Neil {\etal} (1997a)]{one97a} O'Neil, K., Bothun, G.D., and Cornell, M. 1997, \aj, 113, 1212 (OBC)
\bibitem[O'Neil {\etal} (1997b)]{one97b} O'Neil, K., Bothun, G.D., Schombert, J., Cornell, M., and Impey, C.D. 1997, \aj, 114, 2448 
\bibitem[Pennington {\etal} (1993)]{pen93} Pennington, R.L., Humphreys, R.M., Odewahn, S.C., Zumach, W., Thurmes, P.M. 1993, \pasp, 105, 521
\bibitem[Press \etal  (1992)]{pre92} Press, W.H., Teukolsky, S.A., Vetterlin, W.T., and Flannery, B.P. 1992, {\it Numerical Recipes in C (Second Edition)} (Cambridge University Press, New York), 627
\bibitem[Roberts and Haynes (1994)]{rob94} Roberts, M.S. and Haynes, M.P. 1993, \araa, 32, 115
\bibitem[Salzer and Norton (1999)]{sal99} Salzer, J.J. and Norton, S.A. 1999, in {\it The Low Surface Brightness Universe (ASP Conference Series Vol. 170)}, (Astronomical Society of the Pacific, San Francisco), 253.
\bibitem[Schlegel, Finkbeiner, \& Davis (1998)]{sch98} Schlegel, D., Finkbeiner, D.P., and Davis, M. 1998, \aj, 500, 525
\bibitem[Schneider {\etal} (1990)]{sch90} Schneider, S.E., Thuan, T.X., Magri, C. and Wadiak, J.E. 1990, \apjs, 72, 245 
\bibitem[Schneider (1996)]{sch96} Schneider, S.E., 1996, in {\it The Minnesota Lectures on Extragalactic Neutral Hydrogen (ASP Conference Series 106)} (Astronomical Society of the Pacific, San Francisco), 323
\bibitem[Schombert {\etal} (1992)]{sch92} Schombert, J.M., Bothun, G.D., Schneider, S.E., and McGaugh, S.S. 1992, \aj, 103, 1107
\bibitem[Schombert {\etal} (1997)]{sch97} Schombert, J.M., Pildis, R.A., and Eder, J.A. 1997, \apjs, 111, 223
\bibitem[Skillman (1996)]{ski96} Skillman, E.D., 1996, in {\it The Minnesota Lectures on Extragalactic Neutral Hydrogen (ASP Conference Series 106)} (Astronomical Society of the Pacific, San Francisco), 208
\bibitem[Sprayberry, Impey, and Irwin (1996)]{spr96} Sprayberry, D., Impey, C.D., and Irwin M.J. 1996, \apj, 463, 535.
\bibitem[Staveley-Smith, Davies, and Kinman (1992)]{sta92} Stavely-Smith, L., Davies, R.D., and Kinman, T.D. 1992, \mnras, 258, 334
\bibitem[Staveley-Smith, Marquarding, Kilborn, \& Webster(2001)]{sta01} Staveley-Smith, L., Marquarding, M., Kilborn, V.A., \& Webster, R.L.\ 2001, {\it Gas and Galaxy Evolution (ASP Conference Proceedings, Vol. 240)}, (Astronomical Society of the Pacific, San Francisco), 427
\bibitem[Taylor {\etal} (1996)]{tay96} Taylor, C.L. Thomas, D.L., Brinks, E., Skillman, E.D. 1996, \apjs, 107, 143
\bibitem[Theureau {\etal} (1998)]{the98} Theureau, G., Bottinelli, L., Coudreau-Durand, N., Gouguenheim, L., Hallet, N., Loulergue, M., Paturel, G., Teerikorpi, P. 1998, \aaps, 130, 333
\bibitem[van der Hulst {\etal} (1993)]{vdH93} van der Hulst, J.M., Skillman, E.D., Smith, T.R., Bothun, G.D., McGaugh, S.S., \& de Blok, W.J.G. 1993, \aj, 106, 548
\bibitem[van Zee (1999)]{vZe99} van Zee, L. 1999, in {\it The Low Surface Brightness Universe (ASP Conference Series Vol. 170)}, (Astronomical Society of the Pacific, San Francisco), 274.
\bibitem[Verheijen and Sancisi (2001)]{ver01} Verheijen, M.A.W. and Sancisi, R. 2001, \aap, 370, 765
\bibitem[Wegner, Haynes, \& Giovanelli (1993)]{weg93} Wegner, G., Haynes, M.P., Giovanelli, R. 1993, \aj, 105, 1251
\bibitem[Weinberg (1972)]{wei72} Weinberg, S. 1972, {\it Gravitation and Cosmology} (New York:Wiley)
\end{thebibliography}
\end{document}